\documentclass{aa}  
\usepackage{natbib}
\bibpunct{(}{)}{;}{a}{}{,} 

\usepackage{graphicx}
\usepackage[varg]{txfonts}
\newcommand{\va}{c_{\mathrm{A}}}
\newcommand{\csn}{c_{\mathrm{n}}}
\newcommand{\csi}{c_{\rm ie}}

\newcommand{\rhoi}{\rho_{\rm i}}
\newcommand{\rhon}{\rho_{\rm n}}

\newcommand{\xii}{\mbox{\boldmath{$\xi$}}}
\newcommand{\omegaA}{\omega_{\rm A}}
\newcommand{\omegac}{\omega_{\rm c}}
\newcommand{\omegat}{\tilde{\omega}}

\newcommand{\pie}{P'}
\newcommand{\nuin}{\nu_{\rm ni}}
\newcommand{\pn}{p_{\rm n}}
\newcommand{\xiie}{\xi_{r,\rm ie}}
\newcommand{\xin}{\xi_{r,\rm n}}
\newcommand{\ki}{k_{\rm ie}}
\newcommand{\kn}{k_{\rm n}}

\newcommand{\piein}{P'_{0}}
\newcommand{\pieex}{P'_{\rm ex}}
\newcommand{\pnin}{p_{\rm n,0}}
\newcommand{\pnex}{p_{\rm n,ex}}

\begin{document}

	\title{Effect of partial ionization on wave propagation\\ in solar magnetic flux tubes}

	\titlerunning{Waves in a two-fluid flux tube}

\author{R. Soler\inst{\ref{uib}}, A. J. D\'iaz\inst{\ref{iac},\ref{ulag}}, J. L. Ballester\inst{\ref{uib}}, \& M. Goossens\inst{\ref{leuven}}}
\offprints{R. Soler}
\institute{ Solar Physics Group, Departament de F\'isica, Universitat de les Illes Balears,
             E-07122 Palma de Mallorca, Spain. \email{roberto.soler@uib.es} \label{uib} 
 \and
 Instituto de Astrof\'isica de Canarias, E-38200 La Laguna, Tenerife, Spain \label{iac} 
\and
Departamento de Astrof\'isica, Universidad de La Laguna, E-38206 La Laguna, Tenerife, Spain\label{ulag}
\and
Centre for Mathematical Plasma Astrophysics, Department of Mathematics, KU Leuven,
              Celestijnenlaan 200B, 3001 Leuven, Belgium.  
 \label{leuven}}

  \abstract
{Observations show that waves are ubiquitous in the solar atmosphere and may play an important role for  plasma heating. The study of waves in the solar corona is usually based on linear ideal magnetohydrodynamics (MHD) for a fully ionized plasma. However, the plasma in the photosphere and the chromosphere is only partially ionized. Here we investigate theoretically the impact of partial ionization on MHD wave propagation in cylindrical flux tubes in the two-fluid model.  We derive the general dispersion relation that takes into account the effects of neutral-ion collisions and the neutral gas pressure. We take the neutral-ion collision frequency as an arbitrary parameter.  Particular results for  transverse kink modes and slow magnetoacoustic modes are shown. We find that the wave frequencies only depend on the properties of the ionized fluid when the neutral-ion collision frequency is much lower that the wave frequency. For high collision frequencies realistic of the solar atmosphere ions and neutrals behave as a single fluid with an effective density corresponding to the sum of densities of both fluids and an effective sound velocity computed as the average of the sound velocities of ions and neutrals. The MHD wave frequencies are modified accordingly. The neutral gas pressure can be neglected when studying transverse kink waves but it has to be taken into account for a consistent description of slow magnetoacoustic waves. The MHD waves are damped due to neutral-ion collisions. The damping is most efficient when the wave frequency and the collision frequency are of the same order of magnitude.  For high collision frequencies slow magnetoacoustic waves are more efficiently damped than transverse kink waves. In addition, we find the presence of cut-offs for certain combinations of parameters that cause the waves to become non-propagating.}

   \keywords{Sun: oscillations ---
                Sun: atmosphere ---
		Sun: magnetic fields ---
		waves ---
		Magnetohydrodynamics (MHD)}

   \maketitle


\section{Introduction}

Observations show that magnetohydrodynamic (MHD) waves are ubiquitous in the solar atmosphere. Waves are routinely observed in the corona \citep[e.g..,][]{tomczyk2007,tomczyk2009,mcintosh2011}, the chromosphere \citep[e.g..,][]{Kukhianidze2006,depontieu2007,zaqarashvili2007,he2009,okamoto2011}, and the photosphere \citep[e.g.,][]{jess2009,fujimura2009}. In addition, waves and oscillations in thin threads of solar prominences have been reported \citep[e.g.,][]{okamoto2007,lin2007,lin2009,ning2009}.  These recent observations have motivated a  number of theoretical works aiming to understand the properties of the observed waves using the MHD theory \citep[e.g.,][]{vandoorsselaere2008,pascoe2010,pascoe2012,TGV,VTG,soler2011flow,soler2011strat,soler2012}. It is believed that the observed waves may play an important role for the heating of the atmospheric plasma \citep[see, e.g.,][]{erdelyi2007,depontieu2007,mcintosh2011,cargill2011}.

The study of waves in the solar corona is usually based on the linear ideal MHD theory for a fully ionized plasma \citep[see][]{priest1984,goedbloed2004}. In this context, the paper by \citet{edwin1983} on wave propagation in a magnetic flux tube has been used as a basic reference for subsequent works in this field \citep[see also the papers by, e.g.,][]{wentzel1979a,cally1986,goossens2009,goossens2012}. The assumption of full ionization is adequate for the coronal plasma. However, the lower temperatures of the chromosphere and the photosphere cause this assumption to become invalid. Therefore, it is necessary to determine the impact of partial ionization on the wave modes discussed in \citet{edwin1983}  in order to perform a realistic theoretical modeling of wave propagation in partially ionized magnetic wave guides. 

There are many papers which have studied the effect of neutral-ion collision on MHD waves \citep[see, e.g.,][]{brag,depontieu2001,khodachenko2004,khodachenko2006,leake2005,forteza2007}. Since the present paper deals with wave propagation in a magnetic cylinder, here we discuss those works that studied waves in structured media. In solar plasmas the typical frequency of the observed waves is  lower than the expected value of the neutral-ion collision frequency. For this reason the single-fluid theory is usually adopted as a first approximation. The single-fluid approximation assumes a strong coupling between ions and neutrals, so that both fluids behave in practice as a single fluid. \citet{soler2009PI} studied wave propagation in a partially ionized flux tube in the single-fluid approximation. The results of \citet{soler2009PI} show that the waves discussed in \citet{edwin1983} are damped due to neutral-ion collisions. Although partial ionization may be relevant for plasma heating \citep[see][]{khomenko2012}, the damping by neutral-ion collisions is in general too weak to explain the observed rapid attenuation. In addition, \citet{soler2009PI} found critical values for the longitudinal wavelength that constrain wave propagation. However, \citet{zaqarashvili2012} have pointed out that these critical wavelengths may not be physical and may be an artifact of the single-fluid approximation. This is so because the single-fluid approximation misses important effects when small length scales and/or high frequencies are involved. In such cases, the general multifluid description is a more suitable approach \citep[see, e.g.,][]{zaqarashvili2011a}. In a multifluid description no restriction is imposed on the relative values of the wave frequency and the neutral-ion collision frequency. In this more general treatment the mathematical complexity of the equations is substantially increased compared to the single-fluid case.

  A particular form of the multifluid theory is the two-fluid theory in which ions and electrons are considered together as an ion-electron fluid, i.e., the plasma, while neutrals form another fluid that interacts with the plasma by means of collisions. This approach was followed by \citet{kumar2003} to study surface wave propagation in a Cartesian magnetic interface. In the study by \citet{kumar2003} the gas pressure of neutrals was neglected, meaning that the dynamics of neutrals was only governed by the friction force with ions. Recently \citet{soler2012} used the two-fluid theory to study resonant Alfv\'en waves in flux tubes.  \citet{soler2012} focused on transverse kink waves, and so they also neglected gas pressure.

 In the present paper we go beyond these previous studies. We derive the general dispersion relation for waves in cylindrical flux tubes in the two-fluid theory. To do so, we consider a consistent description of the neutral fluid dynamics that includes the effect of neutral gas pressure. This is necessary for a realistic description of magnetoacoustic waves. The use of cylindrical geometry and the consideration of neutral pressure are two significant improvements with respect to the Cartesian model of \citet{kumar2003}. The dispersion relation derived here is the two-fluid generalization of the well-known dispersion relation of \citet{edwin1983} for a fully ionized single-fluid plasma. The collision frequency between ions and neutrals and the ionization degree are two important parameters of the model. The impact of these two parameters on the waves discussed  by \citet{edwin1983} in the fully ionized case is investigated.

This paper is organized as follows. Section~\ref{sec:model} contains a description of the equilibrium configuration and the basic equations. In Section~\ref{sec:normal} we follow a normal mode analysis and derive the general dispersion relation for the wave modes. The case in which neutral gas pressure is neglected is explored in Section~\ref{sec:nopres}, both analytically and numerically. We compare our results with the previous findings of \citet{kumar2003}. Then, we incorporate the effect of neutral gas pressure in Section~\ref{sec:slow}. Finally, we discuss the implications of our results and give our main conclusions in Section~\ref{sec:conc}.

\section{Model and two-fluid equations}
\label{sec:model}

We study  waves in a partially ionized medium composed of ions, electrons, and neutrals. We use the two-fluid theory in which ions and electrons are considered together as an ion-electron fluid, while neutrals form another fluid that interacts with the ion-electron fluid by means of collisions \citep[see, e.g.,][]{zaqarashvili2011a, soler2012,soler2012KHI,diaz2012}.  In all the following expressions, the subscripts `ie' and `n' refer to ion-electrons and neutrals, respectively.

We assume a cylindrically symmetric equilibrium and, for convenience, we use cylindrical coordinates, namely $r$, $\varphi$, and $z$ for the radial, azimuthal, and longitudinal coordinates. The equilibrium is made of a straight magnetic flux tube of radius $R$ embedded in an unbounded environment. The equilibrium magnetic field is straight and constant along the $z$-direction, namely ${\bf B}=B\, \hat{z}$. Gravity is ignored. As a consequence of the force balance condition, the gradients of the equilibrium gas pressures of the different species are zero. We also assume a static equilibrium so that there are no equilibrium flows. We restrict ourselves to the study of linear perturbations superimposed on the equilibrium state. Hence, the governing equations are linearized \citep[see the general expressions in, e.g.,][]{zaqarashvili2011a}. We adopt some additional simplifications that enable us to tackle the problem of wave propagation analytically. We neglect collisions of electrons with neutrals because of the low momentum of electrons. We also drop from the equations the nonadiabatic mechanisms and the magnetic diffusion terms since our purpose here is to determine the impact of neutral-ion collisions only. Thus, the set of coupled differential equations governing linear perturbations from the equilibrium state are 
\begin{eqnarray}
\rho_{\rm i}  \frac{\partial {\bf v}_{\rm i}}{\partial t}  &=& - \nabla p_{\rm ie} + \frac{1}{\mu} \left( \nabla \times {\bf b} \right) \times {\bf B}  -  \rho_{\rm n} \nu_{\rm ni}\left(   {\bf v}_{\rm i} - {\bf v}_{\rm n} \right), \label{eq:momlinion} \\
\rho_{\rm n} \frac{\partial {\bf v}_{\rm n}}{\partial t}  &=& - \nabla p_{\rm n} -  \rho_{\rm n} \nu_{\rm ni} \left( {\bf v}_{\rm n} - {\bf v}_{\rm i} \right),  \label{eq:momlinneu} \\
\frac{\partial {\bf b}}{\partial t} &=& \nabla \times \left( {\bf v}_{\rm i} \times {\bf B} \right),  \label{eq:inductionlin} \\
\frac{\partial p_{\rm ie}}{\partial t}  &=&   - \gamma P_{\rm ie} \nabla \cdot  {\bf v}_{\rm i},  \label{eq:presslinion} \\
\frac{\partial p_{\rm n}}{\partial t}  &=& - \gamma P_{\rm n} \nabla \cdot  {\bf v}_{\rm n},  \label{eq:presslin} 
\end{eqnarray}
where ${\bf v}_{\rm i}$ and ${\bf v}_{\rm n}$ are the velocities of ions and neutrals, respectively, $p_{\rm ie}$ and $p_{\rm n}$ are the pressure perturbations of ions-electrons and neutrals, respectively, ${\bf b}$ is  the magnetic field perturbation,  $\rho_{\rm i}$ and $\rho_{\rm n}$ are the equilibrium densities of ions and neutrals, respectively, $P_{\rm ie}$ and $P_{\rm n}$ are the equilibrium gas pressures of ions-electrons and neutrals, respectively, $\mu$ is the magnetic permittivity, $\gamma$ is the adiabatic index, and  $\nu_{\rm ni}$ is the neutral-ion collision frequency. For our following analysis it is useful to define the ionization degree as $\chi = \rho_{\rm n} / \rho_{\rm i}$. This parameter ranges from $\chi = 0$ for a fully ionized plasma to $\chi \to \infty$ for a neutral gas. Also for convenience, in the following expressions we replace the velocities of ions and neutrals by their corresponding Lagrangian displacements, $\xii_{\rm i}$ and $\xii_{\rm n}$, given by
\begin{equation}
 {\bf v}_{\rm i} = \frac{\partial \xii_{\rm i}}{\partial t}, \qquad {\bf v}_{\rm n} = \frac{\partial \xii_{\rm n}}{\partial t}.
\end{equation}

\section{Normal modes}
\label{sec:normal}

We assume that the equilibrium densities $\rhoi$ and $\rhon$ are functions of $r$ alone, so that the equilibrium is  uniform in both azimuthal and longitudinal directions. The temperatures of ion-electrons and neutrals vary  with $r$ accordingly to keep constant 	the equilibrium gas pressures of both fluids. Hence we can write the perturbed quantities proportional to $\exp \left( i m \varphi + i k_z z  \right)$, where $m$ and $k_z$ are the azimuthal and longitudinal wavenumbers, respectively.  We express the temporal dependence of the perturbations as $\exp \left( - i \omega t \right)$, with $\omega$ the frequency. Now we combine Equations~(\ref{eq:momlinion})--(\ref{eq:presslin}) and arrive at four coupled equations for the radial components of the Lagrangian displacements of ions and neutrals, $\xiie$ and $\xin$ respectively, the ion-electron total pressure Eulerian perturbation, $\pie = p_{\rm ie} + B b_z/\mu$, and the neutral pressure Eulerian perturbation, $\pn$, namely
\begin{eqnarray}
\frac{\partial \pie}{\partial r} &=& \rhoi \left( \omega^2 - \omegaA^2 + i \chi \nuin \omega \right)\xiie \nonumber \\ &-& i \rhon \nuin \omega \xin, \label{eq:pie} \\
\frac{\partial \pn}{\partial r} &=& - i \rhon \nuin \omega \xiie + \rhon \omega \left( \omega + i \nuin \right) \xin, \label{eq:pn} \\
\mathcal{D} \frac{1}{r} \frac{\partial \left( r \xiie \right) }{\partial r} &=& \mathcal{C}_1 \pie + i \frac{\nuin}{\omega + i \nuin}  \mathcal{C}_2 \pn, \label{eq:xiie} \\
\rhon \csn^2 \mathcal{D} \frac{1}{r} \frac{\partial \left( r \xin \right) }{\partial r} &=& i \frac{\nuin}{\omega + i \nuin} \rhon \csn^2  \mathcal{C}_2 \pie + \mathcal{C}_3 \pn, \label{eq:xin}
\end{eqnarray}
with the coefficients $\mathcal{D}$, $\mathcal{C}_1$, $\mathcal{C}_2$, and $\mathcal{C}_3$ defined as
\begin{eqnarray}
\mathcal{D} &=& \rhoi \left( \va^2 + \csi^2  \right) \left( \omegat^2 - \omegaA^2 \right) \left( \omegat^2 - \omegac^2 \right), \\
\mathcal{C}_1 &=& \frac{m^2}{r^2}\left( \va^2 + \csi^2  \right) \left( \omegat^2 - \omegac^2 \right)  \nonumber\\
&-&  \left( \omegat^2 - \omegaA^2 \right) \left( \omegat^2 - k_z^2 \csi^2 \right), \\
\mathcal{C}_2 &=&k_z^2 \csi^2 \left( \omegat^2 - \omegaA^2 \right) + \frac{m^2}{r^2}\left( \va^2 + \csi^2  \right) \left( \omegat^2 - \omegac^2 \right) , \\
\mathcal{C}_3 &=& - \left\{ \frac{\nuin^2}{\left( \omega+i\nuin \right)^2} \rhon \csn^2  \left[ \mathcal{C}_2  + \omegaA^2 \left( \omegat^2 - \omegaA^2 \right)  \right] \right. \nonumber \\
 &+& \left. \frac{\mathcal{D}}{\omega \left( \omega + i \nuin \right)} \left[ \omega \left( \omega + i \nuin \right)-\csn^2 \left( \frac{m^2}{r^2} + k_z^2 \right)\right] \right\},
\end{eqnarray}
where $\omegat^2$, $\omegaA^2$, and $\omegac^2$ are the squares of the modified frequency, the Alfv\'en frequency, and the cusp frequency given by
\begin{eqnarray}
\omegat^2 &=& \omega^2 \left( 1 + \frac{i \chi \nuin}{\omega + i \nuin} \right), \label{eq:omegat} \\
\omegaA^2 &=& k_z^2 \va^2, \\
\omegac^2 &=& k_z^2 \frac{\va^2 \csi^2}{\va^2 + \csi^2 }.
\end{eqnarray}
In addition, $\va^2$ is the square of the Alfv\'en velocity and $\csi^2$ and $\csn^2$ are the squares of the sound velocities of ion-electrons and neutrals, respectively, computed as
\begin{equation}
\va^2 = \frac{B^2}{\mu \rhoi}, \qquad \csi^2 = \frac{\gamma P_{\rm ie}}{\rhoi}, \qquad \csn^2 = \frac{\gamma P_{\rm n}}{\rhon}
\end{equation}

Equations~(\ref{eq:pie})--(\ref{eq:xin})  are  valid for all density profiles in the radial direction. For $\csi = \csn = 0$ and $\pn = 0$ they revert to the equations discussed in \citet{soler2012}. Equations~(\ref{eq:pie})--(\ref{eq:xin}) are singular when $\mathcal{D} = 0$. The positions of the singularities are mobile and depend on the frequency, $\omega$. The whole set of frequencies satisfying $\mathcal{D} = 0$ at some $r$ form two continua of frequencies called the Alfv\'en and cusp (or slow) continua \citep[see, e.g.,][]{appert1974}. A wave whose frequency is within any of the two continua is damped by resonant absorption \citep[see, e.g.,][]{goossens2011}.  Resonantly damped waves in partially ionized magnetic cylinders were studied by \citet{soler2009PIRA} in the single-fluid approximation and by \citet{soler2012} in the two-fluid formalism. 

\subsection{Piece-wise constant equilibrium}

In the present paper we are not interested in studying the resonant behavior of the waves. The interested reader is referred to \citet{soler2009PIRA,soler2012} for studies of resonant waves in partially ionized plasmas. Instead, here we focus on the modification of the wave frequencies due to neutral-ion collisions. For this reason, we avoid the presence of the resonances by choosing both $\rhoi$ and $\rhon$ to be piece-wise constant functions, namely
\begin{equation}
\rhoi(r) = \left\{ \begin{array}{lll}
\rho_{\rm i,0} & \textrm{if}& r \leq R, \\
\rho_{\rm i,ex} & \textrm{if}& r > R, 
\end{array}
\right. \qquad
\rhon(r) = \left\{ \begin{array}{lll}
\rho_{\rm n,0} & \textrm{if}& r \leq R, \\
\rho_{\rm n,ex} & \textrm{if}& r > R, 
\end{array}
\right. 
\end{equation}
where $R$ is the radius of the cylinder with constants densities $\rho_{\rm i,0}$ and $\rho_{\rm n,0}$ embedded in an external environment with constants densities $\rho_{\rm i,ex}$ and $\rho_{\rm n,ex}$. This is the density profile adopted by \citet{edwin1983} in the fully ionized case. Hereafter the indices `0' and `ex' represent the internal and external media, respectively. Note that since the magnetic field and equilibrium pressure of both fluids are constant, the rest of equilibrium quantities are piece-wise constants as well. For simplicity we also take $\nuin$ as a constant free parameter.

For a piece-wise constant equilibrium we can combine Equations~(\ref{eq:pie}) and (\ref{eq:xin}) to eliminate $\xiie$ and $\xin$ and obtain two coupled equations involving $\pie$ and $\pn$ only, namely
\begin{eqnarray}
\frac{\partial^2 \pie}{\partial r^2} + \frac{1}{r} \frac{\partial \pie}{\partial r} + \left( \ki^2 -  \frac{m^2}{r^2} \right) \pie &=& q_{\rm 1}^2 \pn, \label{eq:bessel1} \\
\frac{\partial^2 \pn}{\partial r^2} + \frac{1}{r} \frac{\partial \pn}{\partial r} + \left( \kn^2 -  \frac{m^2}{r^2} \right) \pn &=& q_{\rm 2}^2 \pie, \label{eq:bessel2}
\end{eqnarray}
with
\begin{eqnarray}
\ki^2 &=& \frac{\left( \omega \left( \omega + i \chi \nuin  \right) - \omegaA^2 \right)}{\left( \va^2 + \csi^2  \right)\left( \omegat^2 - \omegac^2 \right)} \nonumber \\ &\times& \left( \omegat^2  - k_z^2 \csi^2 \frac{\omegat^2 - \omegaA^2}{\omega(\omega+i\chi\nuin)-\omegaA^2}\right), \\
\kn^2 &=&  \frac{\omega \left( \omega + i \nuin \right)-k_z^2 \csn^2 }{\csn^2} \nonumber \\ 
&+& \frac{\chi \nuin^2}{\omega + i \nuin} \frac{\omega \omegaA^2}{\left( \va^2 + \csi^2  \right)\left( \omegat^2 - \omegac^2 \right)}, \\
q_{\rm 1}^2 &=& i \frac{\nuin}{\omega + i \nuin} \left[  \frac{\omega \left( \omega + i \nuin \right)-k_z^2 \csn^2 }{\csn^2} + \frac{k_z^2 \csi^2 \left( \omegat^2 - \omegaA^2 \right)}{\left( \va^2 + \csi^2  \right)\left( \omegat^2 - \omegac^2 \right)}  \right. \nonumber \\
&+& \left. \frac{\chi \nuin^2}{\omega+i\nuin} \frac{\omegaA^2 \omega}{\left( \va^2 + \csi^2  \right)\left( \omegat^2 - \omegac^2 \right)} \right], \\
q_{\rm 2}^2 &=& i \frac{\chi \nuin \omega \omegat^2}{\left( \va^2 + \csi^2  \right)\left( \omegat^2 - \omegac^2 \right)}.
\end{eqnarray}
Equations~(\ref{eq:bessel1}) and (\ref{eq:bessel2}) are the basic equations of this investigation. They are two coupled Bessel-type differential equations representing the coupled behavior of ion-electrons and neutrals due to neutral-ion collisions. Note that the terms with $q_1$ and $q_2$ are the ones that couple the equations. These terms vanish in both fully ionized and fully neutral cases. We need to solve Equations~(\ref{eq:bessel1}) and (\ref{eq:bessel2}) to find the general dispersion relation. 

For later use we also give the expressions of $\xiie$ and $\xin$ in terms of $\pie$ and $\pn$, namely
\begin{eqnarray}
\xiie &=& \frac{1}{\rhoi \left( \omegat^2 - \omegaA^2 \right)} \left( \frac{\partial \pie}{\partial r} + i \frac{\nuin}{\omega + i \nuin} \frac{\partial \pn}{\partial r} \right), \\
\xin &=& \left( \frac{1}{\rhon \omega \left(\omega + i \nuin\right)}- \frac{\nuin^2}{\left(\omega + i \nuin\right)^2} \frac{1}{\rhoi \left( \omegat^2 - \omegaA^2 \right)}\right) \frac{\partial \pn}{\partial r}  \nonumber \\
&+& i \frac{\nuin}{\omega + i \nuin} \frac{1}{\rhoi \left( \omegat^2 - \omegaA^2 \right)} \frac{\partial \pie}{\partial r}.
\end{eqnarray}
Note that for $\nuin = 0$ ions and neutrals are decoupled and the magnetic field has no influence on the dynamics of neutrals.

\subsection{Strong thermal coupling and characteristic velocities}

Up to here, no restriction on the values of the  characteristic Alfv\'en and sound velocities  has been made. As a result of the pressure balance condition at $r=R$, the six characteristic velocities in the equilibrium  are related by
\begin{equation}
 \rho_{\rm i,0} \left( \frac{\gamma}{2}c_{\rm A,0}^2 + c_{\rm ie,0}^2 + \chi c_{\rm n,0}^2  \right) =  \rho_{\rm i,ex} \left( \frac{\gamma}{2}c_{\rm A,ex}^2 + c_{\rm ie,ex}^2 + \chi c_{\rm n,ex}^2  \right).
\end{equation}
 Now for simplicity we assume a strong thermal coupling between the ionized and neutral fluids, so that both fluids have locally the same temperature. However, the temperature can still be different in the internal and external media. Using the ideal gas law for both species, this requirement has the consequence that the local sound velocity of the ionized and neutral fluids are related by $\csi^2 = 2\csn^2$. Hence we can define an effective sound velocity of the whole plasma, $c_{\rm s,eff}$, as 
\begin{equation}
c_{\rm s,eff}^2 = \frac{\csi^2 + \chi \csn^2}{1+\chi} = \frac{2+\chi}{2(1+\chi)}\csi^2, \label{eq:effcs}
\end{equation} 
 so that the parameter space of characteristic velocities can be reduced from six to four different velocities, namely $c_{\rm A,0}$, $c_{\rm s,eff,0}$, $c_{\rm A, ex}$, and $c_{\rm s, eff,ex}$.

\subsection{Coupled solutions}

 We continue the mathematical study of the normal modes.  We look for solutions to the coupled Equations~(\ref{eq:bessel1}) and (\ref{eq:bessel2}). In the internal medium, i.e., $r \leq R$, physical solutions imply that both $\pie$ and $\pn$ are regular at $r=0$. The general solutions of $\pie$ and $\pn$ satisfying this condition in the internal medium are
\begin{eqnarray}
\piein &=& A_1 J_m \left( k_{1,0} r \right) + A_2 J_m \left( k_{2,0} r \right), \\
\pnin &=& - A_1 \frac{q_{\rm 2,0}^2}{k_{1,0}^2-k_{\rm n,0}^2} J_m \left( k_{1,0} r \right)\nonumber \\ &-& A_2 \frac{q_{\rm 2,0}^2}{k_{2,0}^2-k_{\rm n,0}^2} J_m \left( k_{2,0} r \right),
\end{eqnarray}
where $J_m$ is the Bessel function of the first kind of order $m$, $A_1$ and $A_2$ are constants, and $k_{1,0}$ and $k_{2,0}$ are the two possible values of the radial wavenumber, $k$, in the internal medium given by the solution of the following equation
\begin{equation}
\left( k^2 - k_{\rm ie,0}^2 \right) \left( k^2 - k_{\rm n,0}^2 \right) - q_{\rm 1,0}^2q_{\rm 2,0}^2 = 0. \label{eq:wavenumber}
\end{equation}
Equivalently, in the external medium, i.e., $r > R$, the solutions of $\pie$ and $\pn$ are of the form
\begin{eqnarray}
\pieex &=& A_3 H^{(1)}_m \left( k_{\rm 1,ex} r \right) + A_4 H^{(1)}_m \left( k_{\rm 2,ex} r \right), \\
\pnex &=& -A_3 \frac{q_{\rm 2,ex}^2}{k_{\rm 1,ex}^2-k_{\rm n,ex}^2} H^{(1)}_m \left( k_{\rm 1,ex} r \right) \nonumber \\
&-& A_4 \frac{q_{\rm 2,ex}^2}{k_{\rm 2,ex}^2-k_{\rm n,ex}^2} H^{(1)}_m \left( k_{\rm 2,ex} r \right),
\end{eqnarray}
where $H^{(1)}_m$ is the Hankel function of the first kind of order $m$, $A_3$ and $A_4$ are constants, and $k_{\rm 1,ex}$ and $k_{\rm 2,ex}$ are the two possible values of the radial wavenumber, $k$, in the external medium also given by the solution of Equation~(\ref{eq:wavenumber}) but now using external values for the parameters. The expressions of $\pie$ and $\pn$ for $r > R$ are written generally in terms of Hankel functions instead of the usual modified Bessel functions, $K_m$, used to represent trapped waves. Thus, our formalism takes into account the possibility of wave leakage, i.e., wave radiation in the external medium. Although we do not explore leaky waves in the present work, the obtained dispersion relation will also be valid for leaky waves.  For ideal trapped waves, the external radial wavenumber is purely imaginary and the function $H^{(1)}_m$ consistently reverts to the modified Bessel function of the second kind, $K_m$, used by \citet{edwin1983}. In the partially ionized case the radial wavenumber is a complex quantity and the use of the function $H^{(1)}_m$ causes us to be very careful when choosing the branch of the external radial wavenumber. To avoid nonphysical energy propagation coming from infinity we need to select the appropriate branch of the external radial wavenumber so that the condition for outgoing waves is satisfied \citep[see details in][]{cally1986,stenuit1999}. 

To understand why there are two different values of the radial wavenumber, $k$, it is instructive to assume a weak coupling between the species, i.e., $\nuin \ll |\omega|$,  so that the quadratic terms in $\nu_{\rm ni}$ can be neglected in Equation~(\ref{eq:wavenumber}). The neglected terms involve the coupling coefficients $q_1$ and $q_2$. Then, the two independent wavenumbers in Equation~(\ref{eq:wavenumber}) simplify to
\begin{equation}
k_1 \approx k_{\rm  ie}, \qquad k_2 \approx k_{\rm n},
\end{equation}
where we have dropped the indices `0' or `ex' because the same result is valid in both internal and external media. The two values of $k$ for weak coupling reduce to the wavenumbers of the ionized and neutral fluids, respectively. For high collision frequencies, these two wavenumbers contain terms with $q_1$ and $q_2$ that couple the two fluids.

\subsection{Dispersion relation}

To find the dispersion relation we match the internal solutions to the external solutions by means of appropriate boundary conditions at $r=R$. The boundary conditions at the interface between two media  in the two-fluid formalism are discussed in \citet{diaz2012}. In the absence of gravity, these boundary conditions reduce to the continuity of $\pie$, $\pn$, $\xiie$, and $\xin$ at $r=R$.  After applying the boundary conditions at $r=R$ we find a system of four algebraic equations for the constants $A_1$, $A_2$, $A_3$, and $A_4$. The condition that the system has a non-trivial solution provides us with the dispersion relation. The dispersion relation is
\begin{equation}
 \mathcal{D}_m \left( \omega, k_z  \right) = 0, \label{eq:reldispergen}
\end{equation}
with the full expression of $\mathcal{D}_m$ given in the Appendix~\ref{app}.  This dispersion relation is the two-fluid generalization of the dispersion relation of \citet{edwin1983} for a fully ionized single-fluid cylindrical flux tube. In the following sections we  discuss the corrections  due to neutral-ion collisions on the wave modes described by \citet{edwin1983} in the fully ionized case. 

 \section{Pressureless neutral fluid}
\label{sec:nopres}

Before exploring the solutions of the general dispersion relation let us consider first the case that neutral pressure is neglected and only the gas pressure of the ionized fluid is taken into account. In the absence of neutral pressure the only force acting on the neutral fluid is the friction force due to neutral-ion collisions.   This is the situation studied by \citet{kumar2003} in planar geometry.  We set $P_{\rm n}= 0$ so that $\csn=0$. After some algebraic manipulations the general dispersion relation simplifies to
\begin{equation}
\frac{k_{0}}{\rho_{\rm i,0} \left( \omegat^2 - \omega_{\rm A,0}^2  \right)} \frac{J_m'(k_{0}R)}{J_m(k_{0}R)} = \frac{k_{\rm ex}}{\rho_{\rm i,ex} \left( \omegat^2 - \omega_{\rm A,ex}^2  \right)} \frac{K^{'}_m(k_{\rm ex}R)}{K_m(k_{\rm ex}R)}, \label{eq:relnopres} 
\end{equation}
with $k_0$ and $k_{\rm ex}$ given by
\begin{eqnarray}
k_0^2 &=& \frac{\left( \omegat^2 - \omega_{\rm A,0}^2 \right) \left( \omegat^2 - k_z^2 c_{\rm ie,0}^2 \right)}{ \left(c_{\rm A,0}^2 + c_{\rm ie,0}^2\right) \left( \omegat^2 - \omega_{\rm c,0}^2 \right)}. \\
k_{\rm ex}^2 &=& - \frac{\left( \omegat^2 - \omega_{\rm A,ex}^2 \right) \left( \omegat^2 - k_z^2 c_{\rm ie,ex}^2 \right)}{ \left(c_{\rm A,ex}^2 + c_{\rm ie,ex}^2\right) \left( \omegat^2 - \omega_{\rm c,ex}^2 \right)}. 
\end{eqnarray}
In  Equation~(\ref{eq:relnopres}) we have used the modified Bessel function of the second kind, $K_m$, instead of the Hankel function of the first kind, $H^{(1)}_m$, in order to write  Equation~(\ref{eq:relnopres}) in the same  form as the dispersion relation of \citet{edwin1983}. Indeed, the comparison between Equation~(\ref{eq:relnopres}) and the dispersion relation of \citet{edwin1983} reveals that both equations are exactly the same if we replace $\omegat^2$ by $\omega^2$ in Equation~(\ref{eq:relnopres}). We can take advantage of this result to easily study the modifications due to neutral-ion collisions of the solutions of the fully ionized case.

\subsection{Analytic approximate study}

\subsubsection{Standing waves}

We focus on standing waves so that we fix the longitudinal wavenumber, $k_z$, to a real value. Due to temporal damping by neutral-ion collisions $\omega$ is complex. i.e., $\omega = \omega_{\rm R} + i \omega_{\rm I}$, where $\omega_{\rm R}$ and $\omega_{\rm I}$ are the real and imaginary parts of the frequency, respectively. Therefore the wave amplitude is damped in time due to the exponential factor $\exp \left( - |\omega_{\rm I}|t \right)$.

We assume that $\omega = \omega_0$ is a trapped solution of the dispersion relation of \citet{edwin1983} in the fully ionized case. Here we only consider trapped waves so that $\omega_0$ is real. For leaky waves see, e.g., \citet{cally1986}. Then $\omegat^2 = \omega_0^2$ is a solution of Equation~(\ref{eq:relnopres}). Using the expression of $\omegat^2$ (Equation~(\ref{eq:omegat})) we expand the Equation $\omegat^2 = \omega_0^2$ as
\begin{equation}
 \omega^3 + i \left( 1+\chi \right) \nuin \omega^2 - \omega_0^2 \omega - i \nuin \omega_0^2=0. \label{eq:relmod}
\end{equation}
The study of the modifications due to neutral-ion collisions of the modes of \citet{edwin1983} reduces to the study of the solutions of Equation~(\ref{eq:relmod}). This study is general and is independent of the particular mode considered since we have not specified what mode $\omega_0$ corresponds to. All the modes are affected in the same way by neutral-ion collisions when neutral pressure is absent.

Equation~(\ref{eq:relmod}) is a cubic equation, hence it has three solutions. To determine the nature of the solutions we perform the change of variable $\omega = -i s$, so that Equation~(\ref{eq:relmod}) becomes a cubic equation in $s$ with all its coefficients real. We compute the discriminant, $\Delta$, of the resulting equation as
\begin{equation}
 \Delta= -\omega_0^2 \left[ 4\left(1+\chi \right)^3 \nuin^4 - \left(  \chi^2 + 20\chi -8\right)\nuin^2 \omega_0^2 +4 \omega_0^4\right],
\end{equation}
The discriminant, $\Delta$, is defined so that (i) Equation~(\ref{eq:relmod}) has one purely imaginary zero and two complex zeros when $\Delta <0$, (ii) Equation~(\ref{eq:relmod}) has a multiple zero and all its zeros are purely imaginary when $\Delta = 0$, and (iii) Equation~(\ref{eq:relmod}) has three distinct purely imaginary zeros when $\Delta >0$. This criterion points out that oscillatory  solutions of Equation~(\ref{eq:relmod}) are only possible when $\Delta <0$.

To determine when oscillatory solutions are not possible we set $\Delta = 0$ and find the corresponding relation between $\omega_0$ and $\nuin$ in terms of $\chi$ as
\begin{equation}
\frac{\nuin}{\omega_0} = \left[\frac{\chi^2+20\chi-8}{8\left( 1+\chi \right)^3} \pm \frac{\chi^{1/2} \left(\chi-8 \right)^{3/2}}{8\left( 1+\chi \right)^3}\right]^{1/2}. \label{eq:rangenoprop}
\end{equation}
Since $\nuin/\omega_0$ must be real, Equation~(\ref{eq:rangenoprop}) imposes a condition on the minimum value of $\chi$ which allows $\Delta= 0$. This minimum value is $\chi = 8$ and the corresponding critical $\nuin/\omega_0$ is $\nuin/\omega_0 = 1/3\sqrt{3} \approx 0.19$. When $\chi < 8$ oscillatory solutions are always possible. When $\chi > 8$ we have to take into account the $+$ and the $-$ signs in Equation~(\ref{eq:rangenoprop}), so that Equation~(\ref{eq:rangenoprop}) defines a range of values of $\nuin/\omega_0$ in which oscillatory solutions are not possible. We call this interval the cut-off region. To our knowledge \citet{kulsrud1969} were the first to report on the existence of a cut-off region of MHD waves in a partially ionized plasma, although \citet{kulsrud1969} only investigated Alfv\'en waves in a homogeneous medium. The existence of cut-off regions for MHD waves in structured media is also deduced from the plots of \citet{kumar2003} but these authors did not explore this phenomenon in detail.  

We look for approximate analytic expressions of the frequency of the oscillatory solutions. Thus we  assume that $\nuin/\omega_0$ is outside the cut-off interval defined by Equation~(\ref{eq:rangenoprop}), meaning that Equation~(\ref{eq:relmod}) has two complex (oscillatory) solutions and one purely imaginary (evanescent) solution. First we focus on the oscillatory solutions. We write $\omega = \omega_{\rm R} + i \omega_{\rm I}$ and insert this expression in Equation~(\ref{eq:relmod}). We assume weak damping and take $|\omega_{\rm I}|\ll |\omega_{\rm R}|$. Hence we neglect terms with $\omega_{\rm I}^2$ and higher powers. It is crucial for the validity of this approximation that $\nuin/\omega_0$ is not within or close to the cut-off region in which $\omega_{\rm R} =0$. After some algebraic manipulations we derive approximate expressions for $\omega_{\rm R}$ and $\omega_{\rm I}$. For simplicity we omit the intermediate steps and give the final expressions, namely
\begin{eqnarray}
 \omega_{\rm R} &\approx &  \omega_0 \sqrt{\frac{\omega_0^2+\left( 1 +\chi \right)\nuin^2}{ \omega_0^2+\left( 1 +\chi \right)^2\nuin^2}}, \label{eq:wr} \\
\omega_{\rm I}&\approx &   -\frac{\chi \nuin}{2\left[  \omega_0^2 +\left( 1+\chi \right)^2 \nuin^2\right]}  \omega_0^2. \label{eq:wi}
\end{eqnarray}
A mode with the same $\omega_{\rm I}$ but with $ \omega_{\rm R}$ of opposite sign is also a solution. When $\nuin = 0$, so that neutral-ion collisions are absent, we find $\omega_{\rm R} = \omega_0$ and $\omega_{\rm I} = 0$. Hence we recover the ideal undamped modes of \citet{edwin1983}. Now these modes are damped due to neutral-ion collisions and the frequency has an imaginary part. In addition, the real part of the frequency is also modified. 

It is instructive to study the limit values of $\nuin$. In the case of low collision frequency, $\nuin^2 \ll \omega_0^2$. Equations~(\ref{eq:wr}) and (\ref{eq:wi}) become
\begin{eqnarray}
 \omega_{\rm R} &\approx &  \omega_0, \label{eq:wrll} \\
\omega_{\rm I}&\approx &   -\frac{\chi \nuin}{2}. \label{eq:will}
\end{eqnarray}
Thus, the real part of the frequency when $\nuin^2 \ll \omega_0^2$ is just the frequency found in the fully ionized case \citep{edwin1983} and it does not depend on the amount of neutrals. The imaginary part of the frequency  is independent of $\omega_0$, meaning that it is the same for all the wave modes. On the other hand, for high collision frequency, $\nuin^2 \gg \omega_0^2$ and Equations~(\ref{eq:wr}) and (\ref{eq:wi}) simplify to
\begin{eqnarray}
 \omega_{\rm R} &\approx &  \frac{\omega_0}{\sqrt{1+\chi}}, \label{eq:wrgg} \\
\omega_{\rm I}&\approx &   -\frac{\chi  \omega_0^2}{2\left( 1+\chi \right)^2 \nuin} . \label{eq:wigg}
\end{eqnarray}
Now the expression of $\omega_{\rm R}$ involves $\chi$ in the denominator, so that the larger the amount of neutrals, the lower $\omega_{\rm R}$ compared to the frequency in the fully ionized case. The frequency is reduced by the factor $1/\sqrt{1+\chi}$ compared to the fully ionized value. This is the same factor found by \citet{kumar2003}  and is equivalent to replace the ion density by the sum of densities of ions and neutrals or, equivalently, to replace $\rhoi$ by $ \left( 1 + \chi \right) \rhoi$.  The imaginary part of the frequency depends now on $\omega_0$, meaning that the various modes have different damping rates.

In addition to the oscillatory solutions, Equation~(\ref{eq:relmod}) has one purely imaginary solution whose approximation is given by
\begin{equation}
 \omega \approx - i \nuin \frac{\omega_0^2+\left( 1 +\chi \right)^2\nuin^2}{\omega_0^2+\left( 1 +\chi \right)\nuin^2}. \label{eq:evanescent}
\end{equation}
The perturbations related to these modes are evanescent in time. There exist a different evanescent solution related to each mode, $\omega_0$, of the fully ionized case.  Purely imaginary solutions were also found by \citet{zaqarashvili2011a} when studying waves in a partially ionized homogeneous medium. These modes are not present in the fully ionized case. These evanescent perturbations may be relevant during the excitation of the waves, since part of the energy used to excite the waves may go to these modes instead of being used to excite the oscillatory solutions. This could be investigated by going beyond the present normal mode analysis and solving the initial value problem. 

\subsubsection{Propagating waves}

Now we turn to propagating waves. In the fully ionized case the study of propagating waves is equivalent to that of standing waves. Here we shall see that neutral-ion collisions break this equivalence and propagating waves are worth being studied separately. Propagating waves were not investigated by \citet{kumar2003}. For propagating waves, we fix the frequency, $\omega$, to a real value and solve the dispersion relation for the complex $k_z$, i.e., $k_z = k_{z,\rm R}+i k_{z,\rm I}$, where $k_{\rm R}$ and $k_{\rm I}$ are the real and imaginary parts of $k_z$, respectively.  Therefore the wave amplitude is damped in space due to the exponential factor $\exp \left( - k_{z,\rm I} z \right)$.  

Equation~(\ref{eq:relnopres}) also holds for propagating waves. We assume that $k_z = k_{z,0}$, with $ k_{z,0}$ real, is a solution of the ideal dispersion relation of \citet{edwin1983}. Then, 
\begin{equation}
k_z^2 = k_{z,0}^2 \frac{\omega + i \left( 1 + \chi \right)\nuin}{\omega + i\nuin}, \label{eq:kz}
\end{equation}
is a solution of Equation~(\ref{eq:relnopres}) following the same argument as  explained in the case of standing waves. We write  $k_z = k_{z,\rm R}+i k_{z,\rm I}$ and insert this expression in Equation~(\ref{eq:kz}). We obtain exact expressions for $k_{z,\rm R}$ and $k_{z,\rm I}$, namely
\begin{eqnarray}
k_{z,\rm R} &=& k_{z,0} \sqrt{\frac{\omega^2+(1+\chi)\nuin^2}{2\left( \omega^2+\nuin^2 \right)}} \nonumber \\ &\times& \left[ 1 \pm \left( 1 + \frac{\chi^2\nuin^2\omega^2}{\left( \omega^2+(1+\chi)\nuin^2 \right)^2} \right)^{1/2}  \right]^{1/2}, \label{eq:ksr2} \\
k_{z,\rm I} &=& \frac{k_{z,0}^2}{2k_{z,\rm R}} \frac{\chi\nuin\omega}{\omega^2+\nuin^2}.  
\end{eqnarray}
The same values of $k_{z,\rm R}$ and $k_{z,\rm I}$ but with opposite signs are also solutions.  In principle, Equation~(\ref{eq:ksr2}) allows two different values of $k_{z,\rm R}$ because of the $\pm$ sign. However the solution with the $-$ sign is not physical since it corresponds to $k_{z,\rm R}$ imaginary, which is an obvious contradiction.  For this reason we discard the solution with the $-$ sign and take the $+$ sign in Equation~(\ref{eq:ksr2}). Now we realize that
\begin{equation}
\frac{\chi^2\nuin^2\omega^2}{\left( \omega^2+(1+\chi)\nuin^2 \right)^2} \ll 1.
\end{equation}
Thus we can approximate $k_{z,\rm R}$ and $k_{z,\rm I}$ as
\begin{eqnarray}
k_{z,\rm R} &\approx& k_{z,0} \sqrt{\frac{\omega^2+(1+\chi)\nuin^2}{\omega^2+\nuin^2 }}, \label{eq:kr} \\
k_{z,\rm I} &\approx& \frac{k_{z,0}}{2} \frac{\chi\nuin\omega}{\sqrt{\left(\omega^2+\nuin^2\right)\left( \omega^2+(1+\chi)\nuin^2 \right)}}.  \label{eq:ki} 
\end{eqnarray}
These expressions are equivalent to Equations~(\ref{eq:wr}) and (\ref{eq:wi}) obtained for standing waves. However, contrary to Equations~(\ref{eq:wr}) and (\ref{eq:wi}), Equations~(\ref{eq:kr}) and (\ref{eq:ki}) are valid for all values of the ratio $\nuin/\omega$. There is no cut-off region for propagating waves because $k_{z,\rm R}$ never vanishes. There are no purely imaginary solutions of $k_z$, i.e., all the solutions have always an oscillatory behavior in $z$. This is an important difference with the case of standing waves.

\subsection{Comparison with numerical results}

Here we numerically solve the dispersion relation (Equation~(\ref{eq:relnopres})). For simplicity we focus on standing waves and show the results for transverse kink ($m=1$) waves only, although we have checked that equivalent results are found in the case of other modes as, e.g., slow magnetoacoustic modes. In the fully ionized case, the frequency of the transverse kink waves in the thin tube (TT) limit, i.e., $k_z R \ll 1$, is $\omega = \omega_{\rm k}$, with $\omega_{\rm k}$  the kink frequency given by
\begin{equation}
 \omega_{\rm k} = k_z \sqrt{\frac{\rho_{\rm i,0}c_{\rm A,0}^2 + \rho_{\rm i,ex}c_{\rm A,ex}^2 }{\rho_{\rm i,0} + \rho_{\rm i,ex}}}. \label{eq:wk}
\end{equation}

\begin{figure}[!t]
\centering
  \includegraphics[width=0.85\columnwidth]{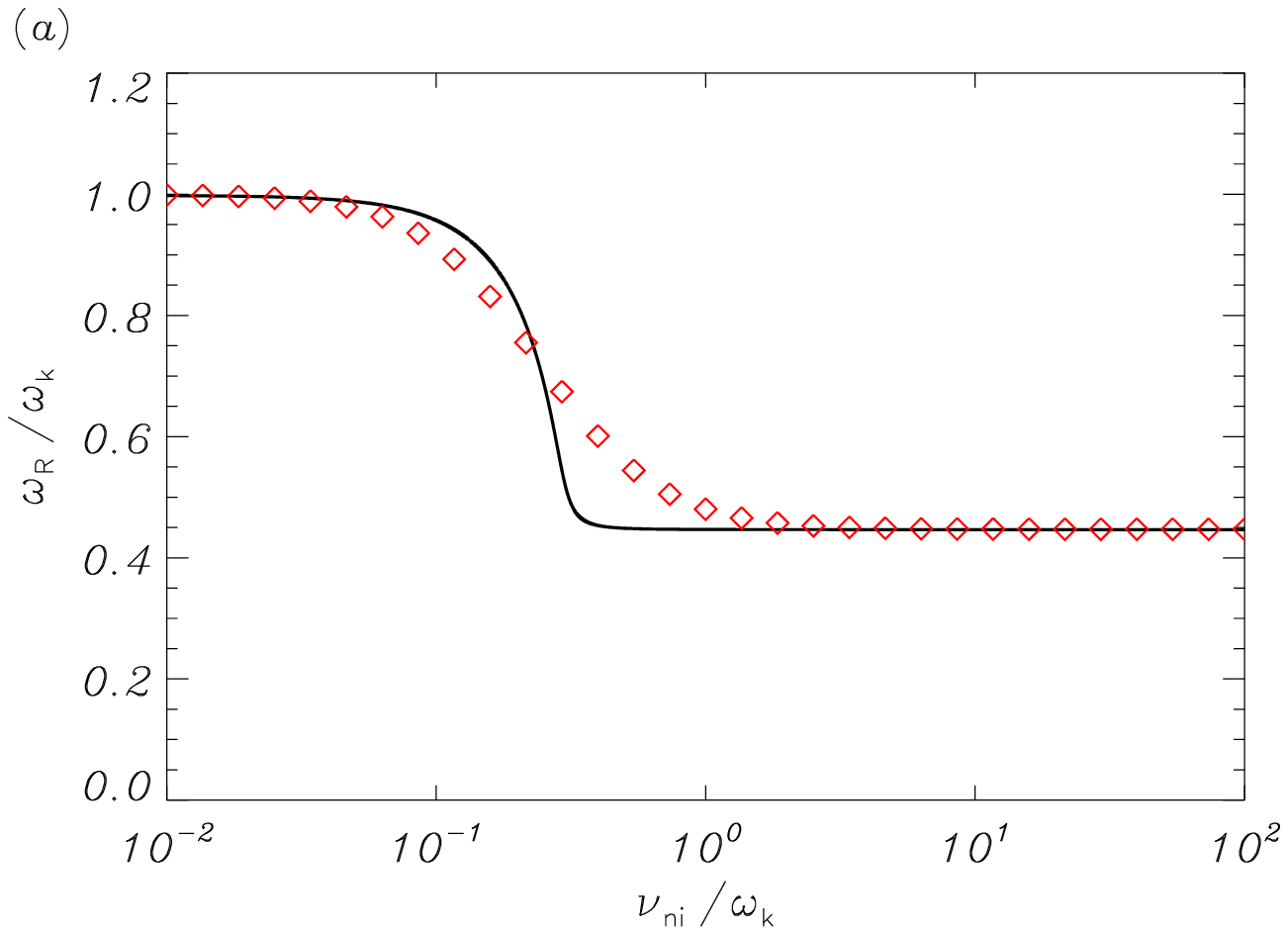}
\includegraphics[width=0.85\columnwidth]{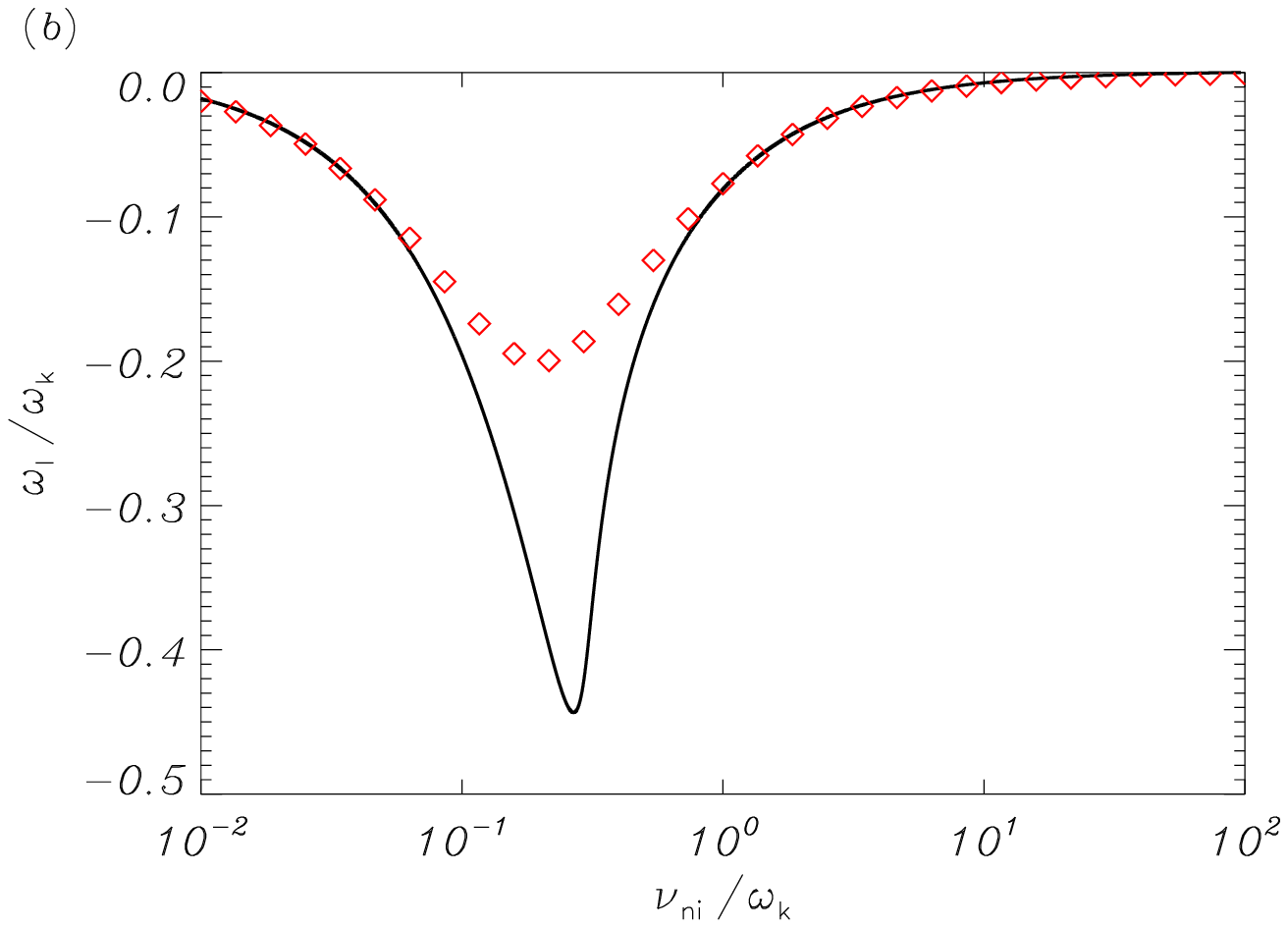}
\caption{ (a) $\omega_{\rm R} / \omega_{\rm k}$ and (b) $\omega_{\rm I} / \omega_{\rm k}$ versus $\nuin / \omega_{\rm k}$ for the transverse kink ($m=1$) mode with $k_z R = 0.1$, $ \rho_{\rm i,0} /\rho_{\rm i,ex} = 3$, $c_{\rm ie,0}/c_{\rm A,0} = 0.2$, and $\chi=4$. The solid line is the result obtained by numerically solving the dispersion relation in the absence of neutral pressure (Equation~(\ref{eq:relnopres})). The symbols are the approximate analytic results in the weak damping approximation (Equations~(\ref{eq:wr}) and (\ref{eq:wi})). \label{fig:kink1}}
\end{figure}

Figure~\ref{fig:kink1} displays the real and imaginary parts of the frequency of the kink mode as functions of the ratio $\nuin / \omega_{\rm k}$ for a particular choice of parameters given in the caption of the figure. We compare the numerical results with the analytic approximations. In Figure~\ref{fig:kink1} we have considered $\chi < 8$ so that no cut-off region is present. Regarding the real part of the frequency (Figure~\ref{fig:kink1}a), we obtain that for $\nuin / \omega_{\rm k} \ll 1$  the results are independent of the ionization degree, with $\omega_{\rm R} \approx \omega_{\rm k}$ as in the fully ionized case. When the ratio $\nuin / \omega_{\rm k}$ increases, plasma and neutrals are more coupled and the ionization degree becomes a relevant parameter, so that $\omega_{\rm R} $ decreases until de value $\omega_{\rm R} \approx \omega_{\rm k}/\sqrt{1+\chi}$ is reached. This behavior is consistent with the analytic Equation~(\ref{eq:wr}) although the details of the transition are not fully captured by the approximation. Regarding the imaginary part of the frequency (Figure~\ref{fig:kink1}b), we find that $\omega_{\rm I}$ tends to zero in the limits  $\nuin / \omega_{\rm k} \ll 1$ and $\nuin / \omega_{\rm k} \gg 1$.  The damping is most efficient when $\nuin$ and $\omega_{\rm k}$ are of the same order of magnitude, approximately. This result is also consistent  with the analytic Equation~(\ref{eq:wi}), although the approximation underestimates the actual damping rate when $\omega_{\rm I}$ is minimal. This is so because Equation~(\ref{eq:wi}) was derived in the weak damping approximation, i.e., $|\omega_{\rm I}|\ll |\omega_{\rm R}|$, while $|\omega_{\rm I}|$ and $|\omega_{\rm R}|$ are of the same order when $\omega_{\rm I}$ is minimal, meaning that the damping is strong.

\begin{figure}[!t]
\centering
  \includegraphics[width=0.85\columnwidth]{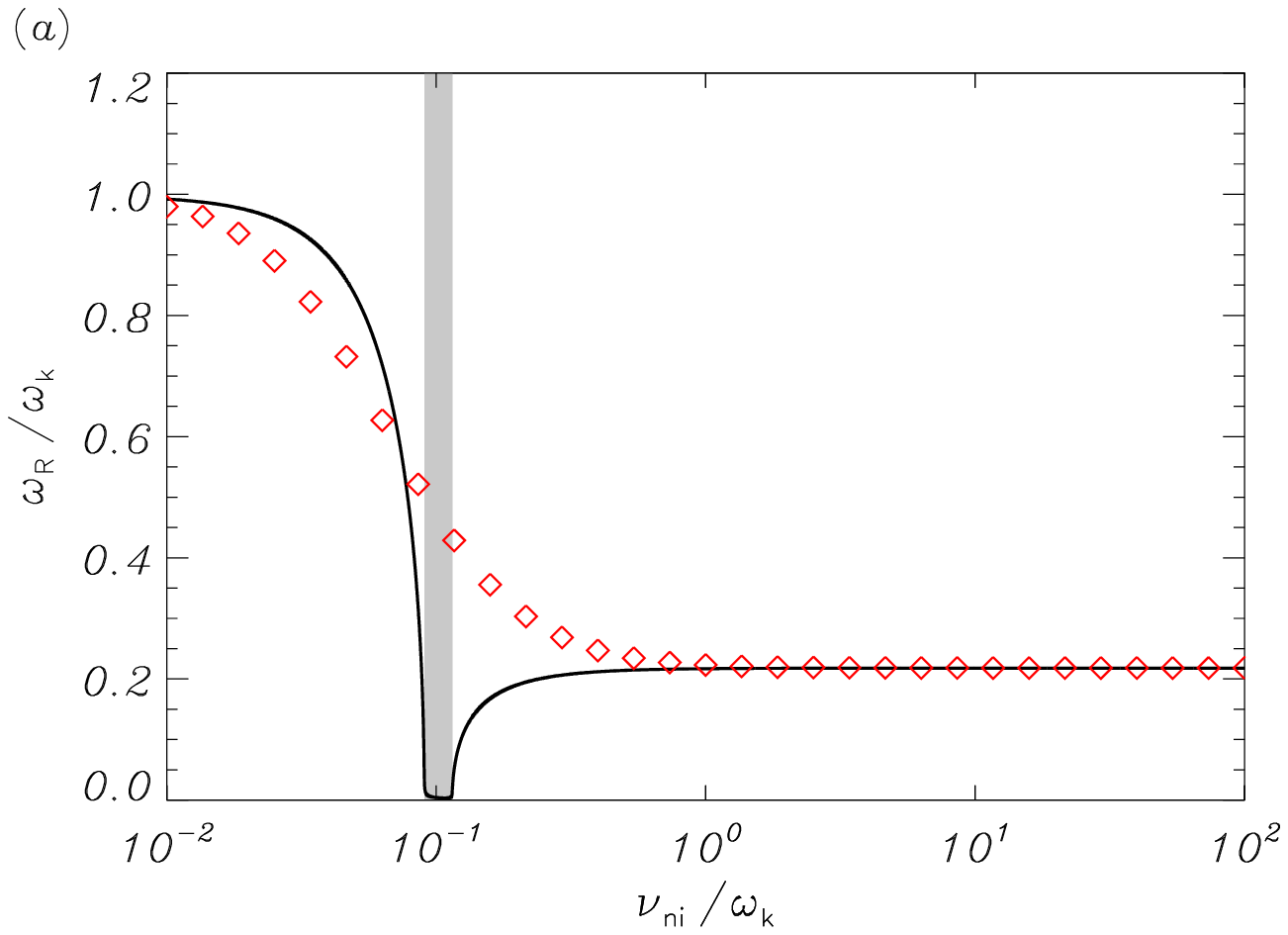}
\includegraphics[width=0.85\columnwidth]{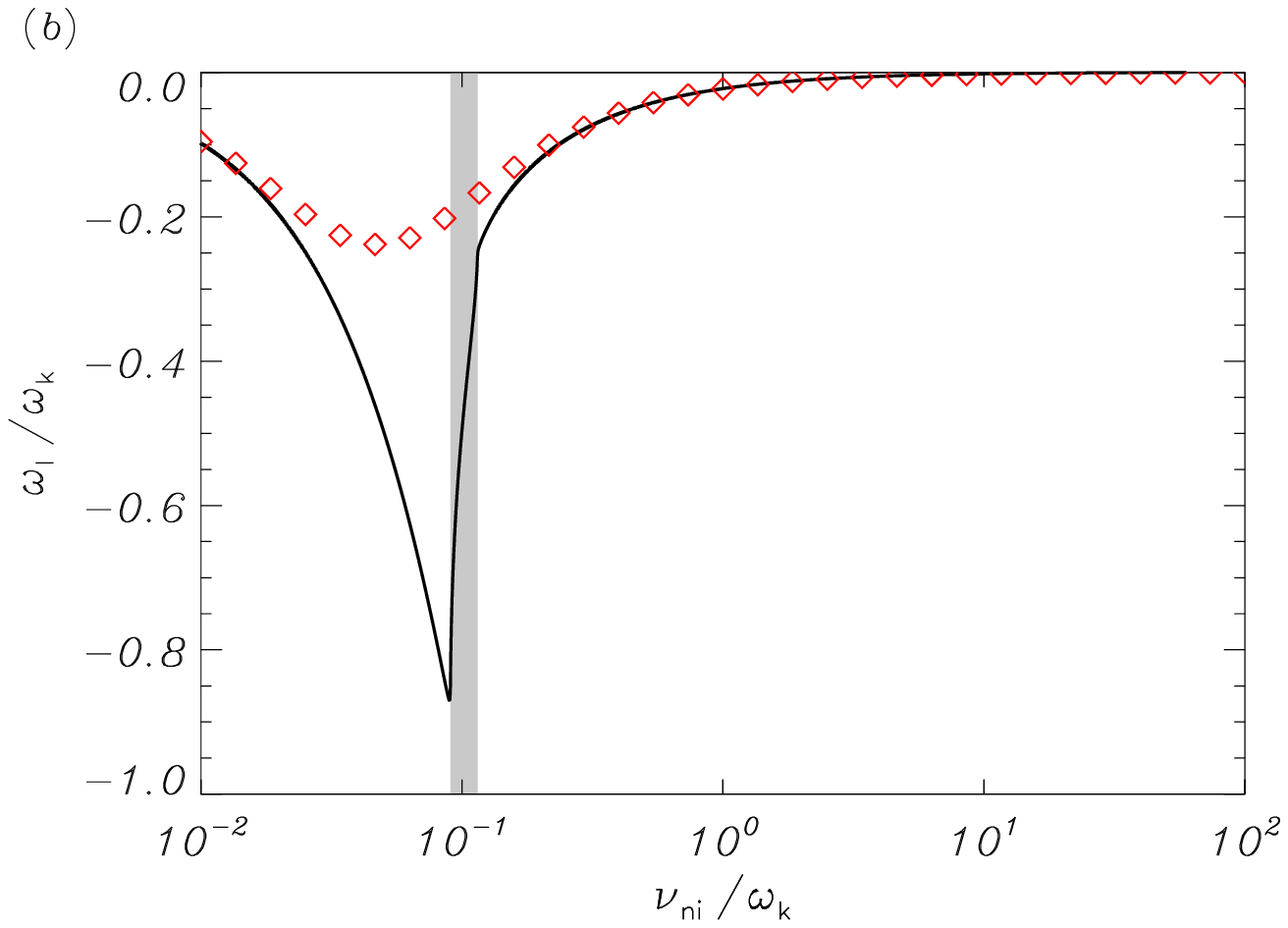}
\caption{ Same as Figure~\ref{fig:kink1} but with $\chi=20$. The shaded area denotes the cut-off region given by Equation~(\ref{eq:rangenoprop}). \label{fig:kink2}}
\end{figure}

Now we take $\chi > 8$ and repeat the previous computations. These results are shown in Figure~\ref{fig:kink2}. We notice the presence of a cut-off region for a certain range of $\nuin / \omega_{\rm k}$. In this cut-off region  $\omega_{\rm R} =0$. The location of the cut-off region agrees very well with the range given by Equation~(\ref{eq:rangenoprop}). This zone exists for relatively small $\nuin / \omega_{\rm k}$. In the solar atmosphere the expected value of the collision frequency \citep[see, e.g.,][Figures~1 and 2]{depontieu2001} is much higher than the  frequency of the observed waves  \citep[e.g.,][]{depontieu2007,okamoto2011}. The minimum value of the neutral-ion collision frequency in the solar atmosphere is of the order of 10~Hz, while the dominant frequency in the observations of chromospheric kink waves by \citet{okamoto2011} is around 22~mHz. Therefore, the cut-off region found here is not in the range of $\nuin / \omega_{\rm k}$ consistent with solar atmospheric parameters and observed wave frequencies. However, in other astrophysical situations the presence of the cut-off region may be relevant \citep[see][]{kulsrud1969}.  As expected, the analytic approximations (Equations~(\ref{eq:wr}) and (\ref{eq:wi})) completely miss the presence of the cut-off region, although they are reasonably good far from the location of the cut-off.

 It is worth mentioning again that  the effect of resonant absorption on the damping of transverse kink waves is absent due to the use of a piece-wise constant density profile \citep{rudermanroberts2002,goossens2002}. \citet{soler2012} showed that damping by resonant absorption is more efficient than damping by neutral-ion collisions when $\nuin / \omega_{\rm k} \gg 1$. Hence, the actual damping of transverse kink waves would be stronger than the damping shown here if resonant absorption were taken into account \citep[see details in][]{soler2012}.

\section{Role of neutral pressure}

\label{sec:slow}

We incorporate the effect of neutral pressure. We anticipate that neutral pressure is relevant for slow magnetoacoustic waves, while its effect on transverse kink waves is minor. \citet{goossens2009,goossens2012} showed that transverse kink waves in flux tubes have the typical properties of surface Alfv\'en (or Alfv\'enic) waves. In thin tubes, i.e., $k_zR\ll 1$, the main restoring force of these waves is magnetic tension and the gas pressure force is negligible. We have computed the kink mode  frequency in the presence of neutral pressure and have found no significant differences with the result in the absence of neutral pressure. Hence, here we focus on the study of slow magnetoacoustic waves and  explore the modification of the slow mode frequency due to neutral-ion collisions when neutral pressure is taken into account. In the fully ionized case \citep{edwin1983},  slow magnetoacoustic waves in the TT limit have frequencies $\omega \approx \omega_{\rm c,0} $, with $\omega_{\rm c,0}$ the internal cusp frequency given by 
\begin{equation}
 \omega_{\rm c,0} =  k_z\sqrt{\frac{c_{\rm A,0}^2 c_{\rm ie,0}^2}{c_{\rm A,0}^2 + c_{\rm ie,0}^2}}.\label{eq:cuspfreq}
\end{equation}
This result is independent of the azimuthal wavenumber, $m$. In a low-$\beta$ plasma, i.e., when the magnetic pressure is much more important than the gas pressure, $\omega_{\rm c,0} \approx k_z c_{\rm ie,0}$. Due to the complexity of the general dispersion relation (Equation~(\ref{eq:reldispergen})) it is not possible to obtain simple analytic approximations of the slow mode frequency. For this reason we perform this investigation in a numerical way. 

\begin{figure}[!t]
\centering
  \includegraphics[width=0.85\columnwidth]{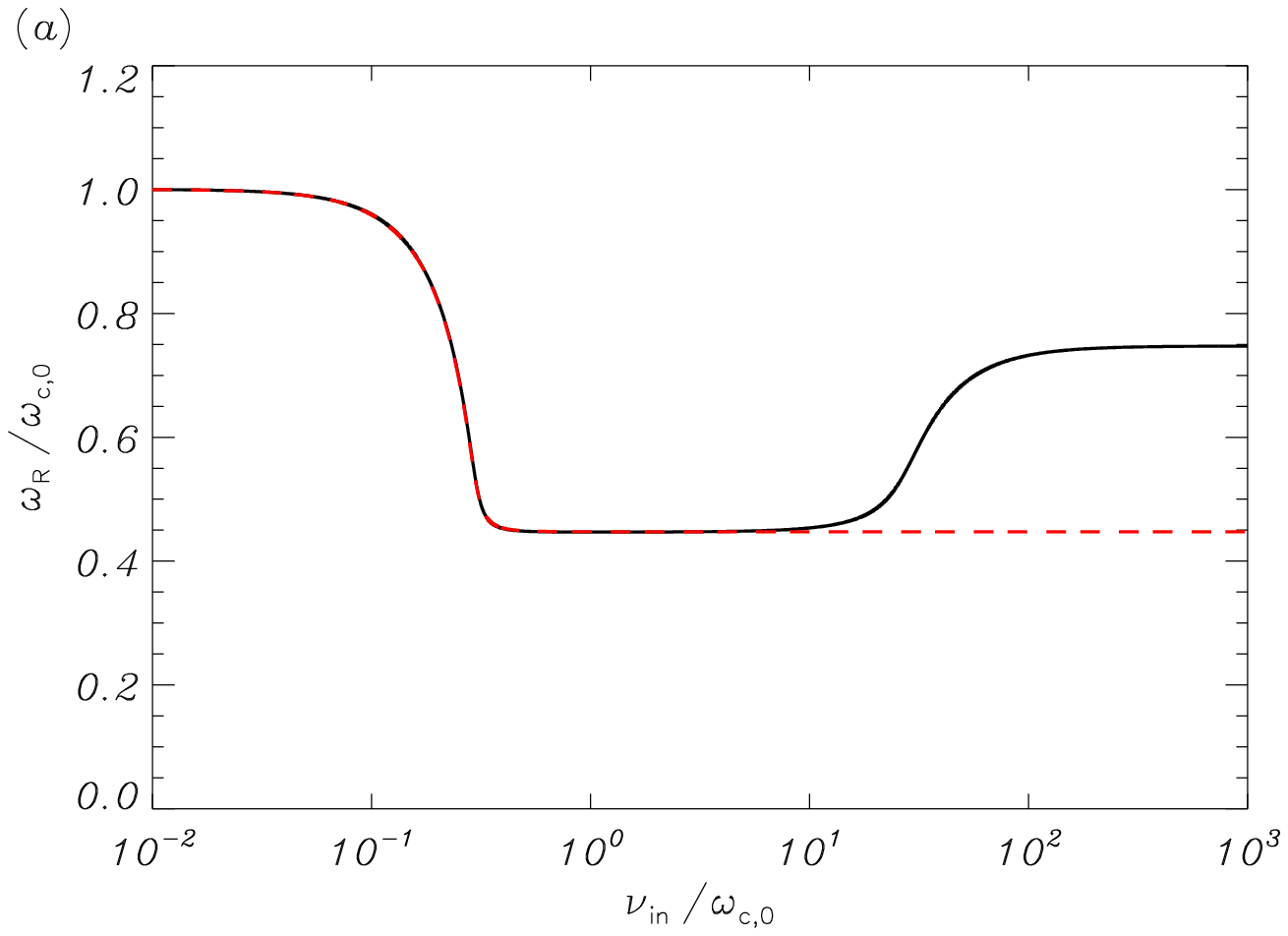}
\includegraphics[width=0.85\columnwidth]{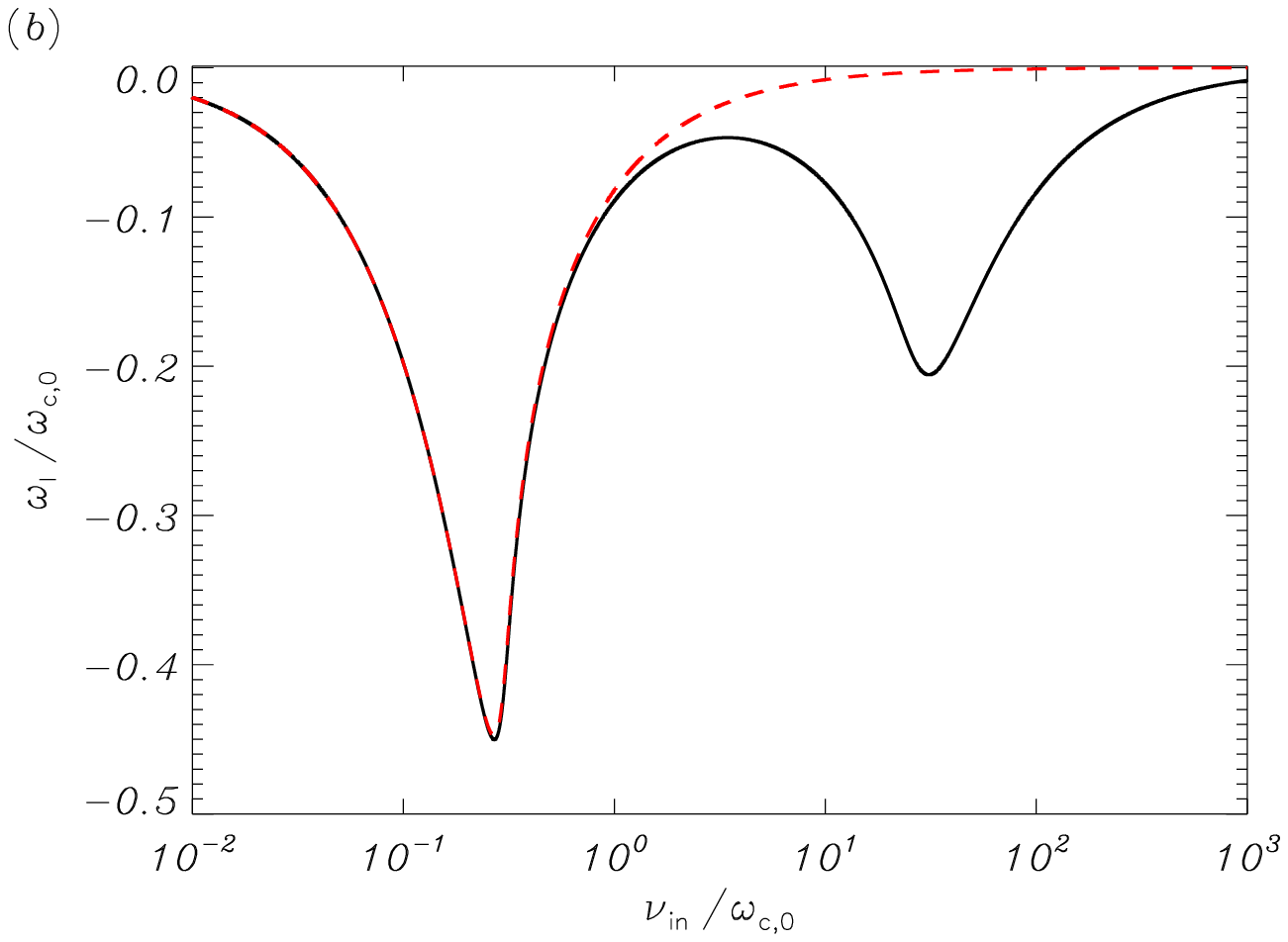}
\caption{ (a) $\omega_{\rm R} / \omega_{\rm c,0}$ and (b) $\omega_{\rm I} / \omega_{\rm c,0}$ versus $\nuin / \omega_{\rm c,0}$ for the slow magnetoacoustic mode with $\chi=4$ and the rest of parameters the same as in Figure~\ref{fig:kink1}. The solid line is the full result taking neutral pressure into account, while the dashed line is the result when neutral pressure is absent. \label{fig:slow1}}
\end{figure}

Figure~\ref{fig:slow1} shows the numerically obtained real and imaginary parts of the slow mode frequency as functions of the ratio $\nuin / \omega_{\rm c,0}$ for $\chi<8$. The rest of parameters are given in the caption of the figure. For comparison, we overplot the results in the absence of neutral pressure which are obtained by solving Equation~(\ref{eq:relnopres}) with the same parameters. First we discuss the behavior of the $\omega_{\rm R}$ (Figure~\ref{fig:slow1}a). When $\nuin / \omega_{\rm c,0}  \ll 1$ we recover the result in the fully ionized case, i.e., $\omega_{\rm R} \approx \omega_{\rm c,0} $. When $\nuin / \omega_{\rm c,0}$ increases, the real part of the slow mode frequency decreases until a plateau is reached at the value $\omega_{\rm R} \approx \omega_{\rm c,0}/\sqrt{1+\chi}$. This is equivalent to perform the replacements $\va^2 \to \va^2/(1+\chi)$ and $\csi^2 \to \csi^2/(1+\chi)$ in Equation~(\ref{eq:cuspfreq}) and is the same behavior as that found in the absence of neutral pressure, so that both results are superimposed in Figure~\ref{fig:slow1}a. However, as the ratio $\nuin / \omega_{\rm c,0}$ continues to increase, the real part of the frequency rises again until a second plateau is finally reached. The presence of this second plateau of $\omega_{\rm R}$ for large collision frequencies owes its existence to the effect of neutral pressure and is absent when neutral pressure is neglected. 

Let us analyze in detail the second plateau of $\omega_{\rm R}$ at high collision frequencies. Since we have no simple analytic expressions for the slow mode frequency  we have performed a parameter study. We have considered various values of $\csi$ and have numerically computed the slow mode $\omega_{\rm R}$ at the second plateau. The result of this parameter study (not shown here for simplicity) points out that the value of  $\omega_{\rm R}$ at the second plateau is, approximately,
 \begin{equation}
  \omega_{\rm R}  \approx \omega_{\rm c,0} \sqrt{\frac{2+\chi}{1+\chi}}\left(2+\frac{\chi c_{\rm ie,0}^2}{c_{\rm A,0}^2 + c_{\rm ie,0}^2}\right)^{-1/2}. \label{eq:wrslow2}
 \end{equation}
This expression can be obtained from Equation~(\ref{eq:cuspfreq}) by performing the replacements $\va^2 \to \va^2/(1+\chi)$ and $\csi^2 \to c^2_{\rm s,eff}$, with the expression of the effective sound velocity, $c_{\rm s,eff}$, given in Equation~(\ref{eq:effcs}). In a low-$\beta$ plasma, $\va^2 \gg \csi^2$ and $\omega_{\rm c,0} \approx k_z c_{\rm ie,0}$ so that Equation~(\ref{eq:wrslow2}) simplifies to
\begin{equation}
\omega_{\rm R}  \approx k_z c_{\rm ie,0} \sqrt{\frac{2+\chi}{2(1+\chi)}} = k_z c_{\rm s,eff,0}.
\end{equation}
Hence, for practical purposes we obtain that at high collision frequencies the sound velocity of the ionized fluid has to be replaced by the effective sound velocity in the expressions of the frequency. The effective sound velocity takes into account the sound velocity of neutrals in addition to the sound velocity of ions. On the contrary, for intermediate collision frequencies it is enough to replace $\csi$ by $\csi/\sqrt{1+\chi}$, so that the sound velocity of neutrals can be ignored.

We turn to the imaginary part of the frequency (Figure~\ref{fig:slow1}b).  As for the real part of the frequency, there are striking differences between the results with and without neutral pressure. The full result shows the presence of two different minima of $\omega_{\rm I}$, whereas only the first minimum is present  in the absence of neutral pressure. The second minimum takes place in the range of high collision frequencies compared to the ideal cusp frequency. As a consequence of the presence of this second minimum for high collision frequencies, the efficiency of neutral-ion collisions for the damping of the slow mode is larger than for transverse kink modes when $\nuin$ approaches realistic values. The stronger damping of the slow modes due to neutral-ion collisions compared to the weak damping of the transverse kink modes was already obtained by \citet{soler2009PI} in the single-fluid approximation.

\begin{figure}[!t]
\centering
  \includegraphics[width=0.85\columnwidth]{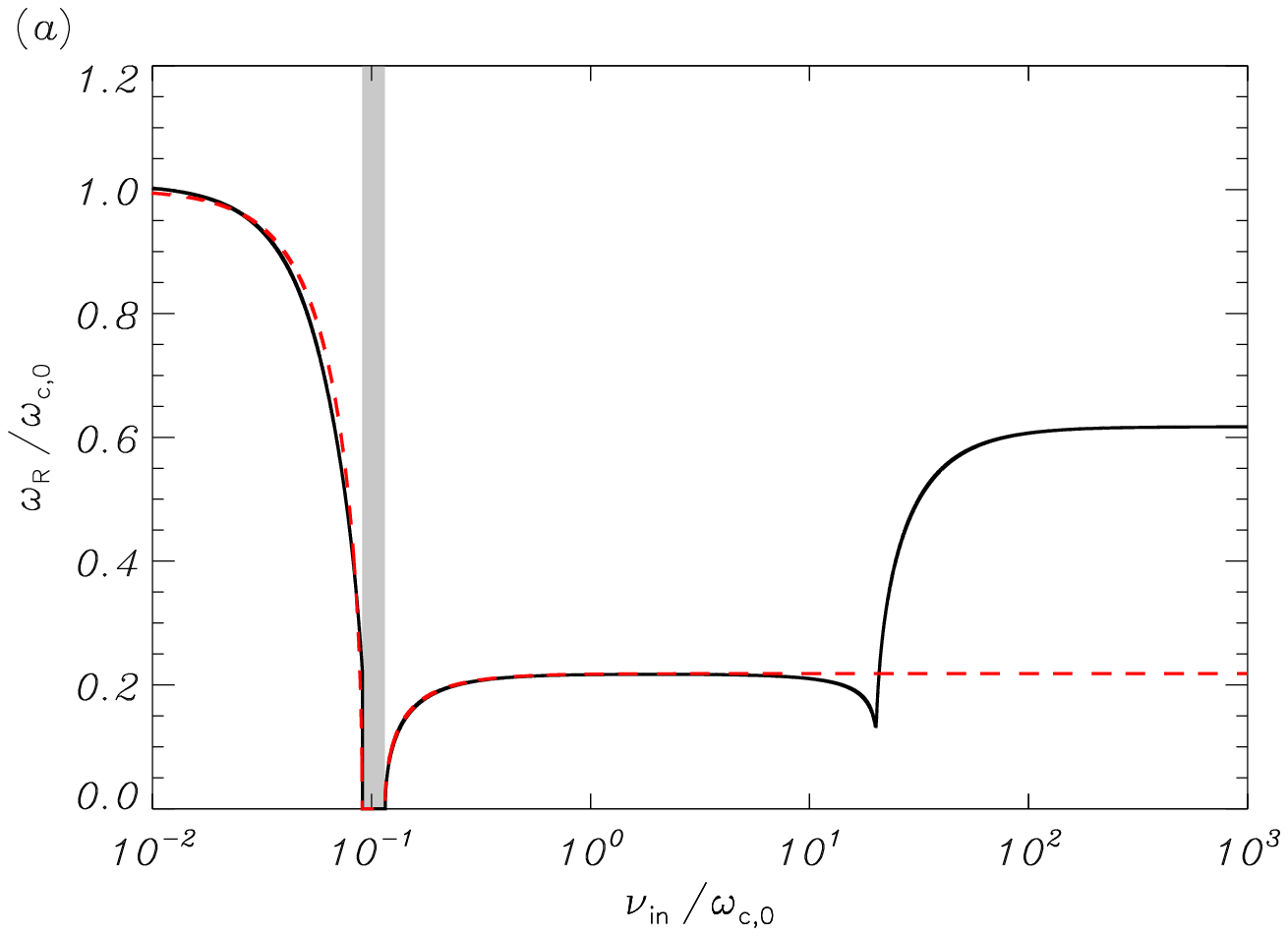}
\includegraphics[width=0.85\columnwidth]{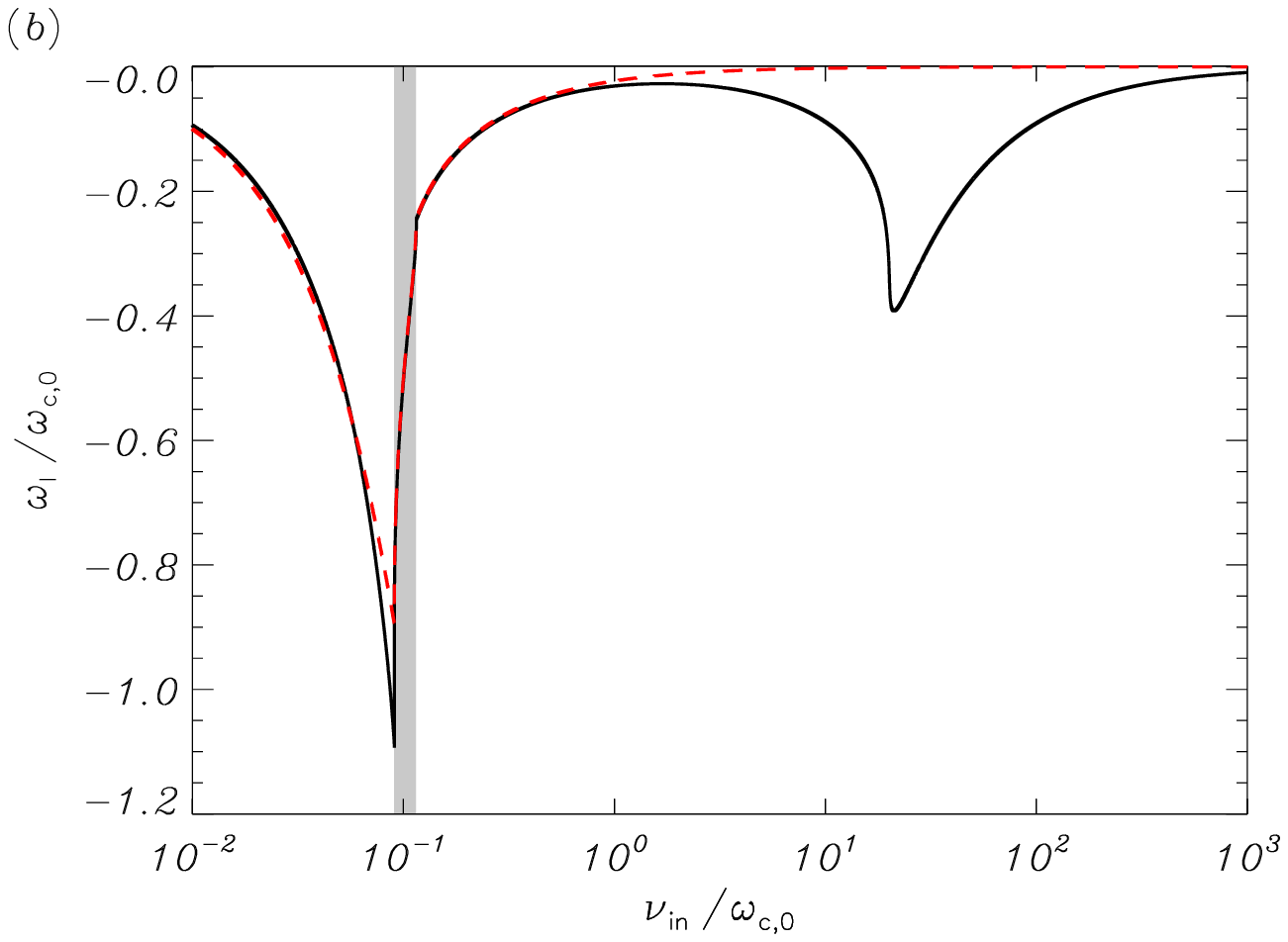}
\caption{Same as Figure~\ref{fig:slow1} but with $\chi=20$. The shaded area denotes the cut-off region given by Equation~(\ref{eq:rangenoprop}).  \label{fig:slow2}}
\end{figure}

\begin{figure}[!t]
\centering
  \includegraphics[width=0.85\columnwidth]{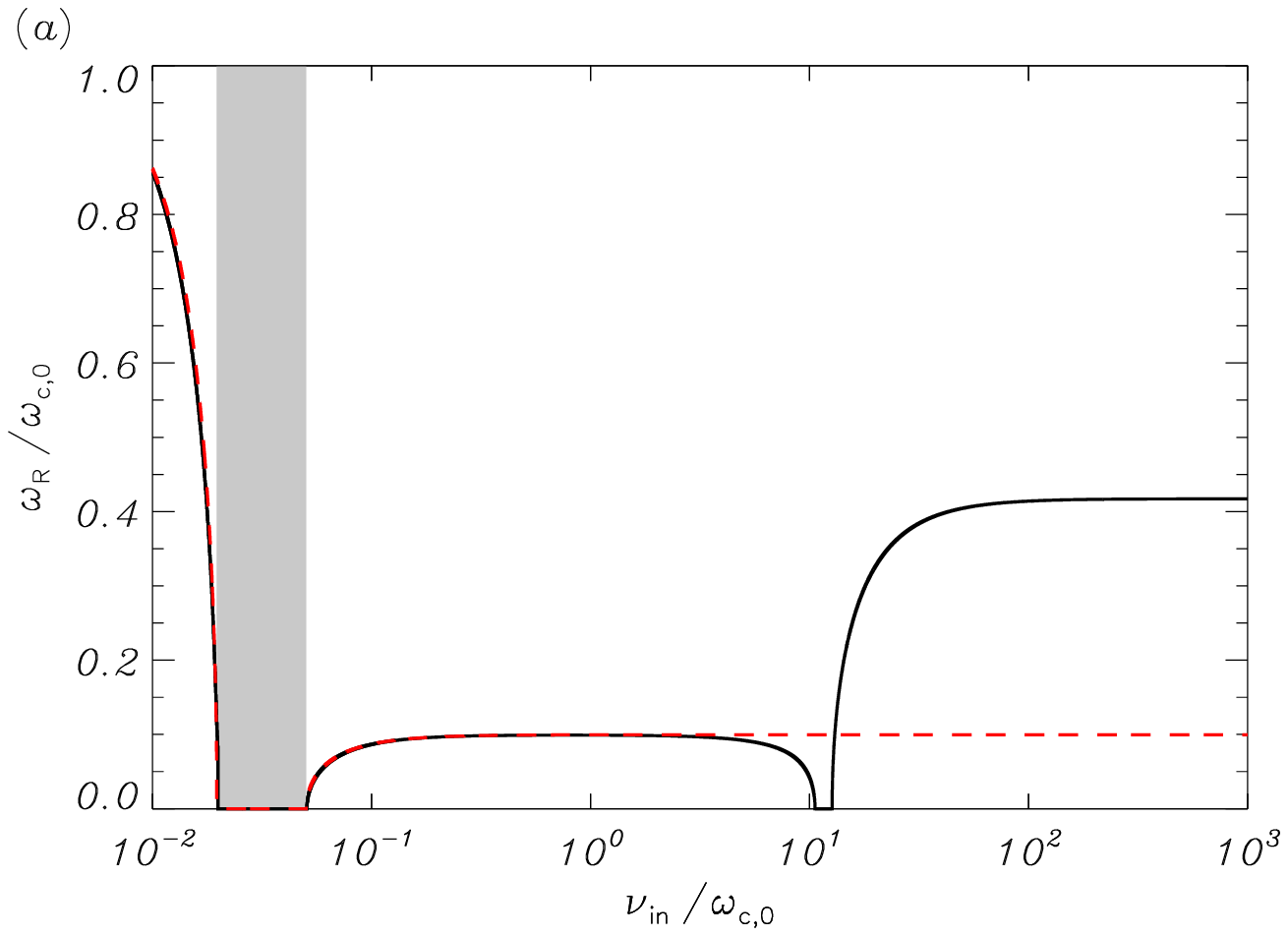}
\includegraphics[width=0.85\columnwidth]{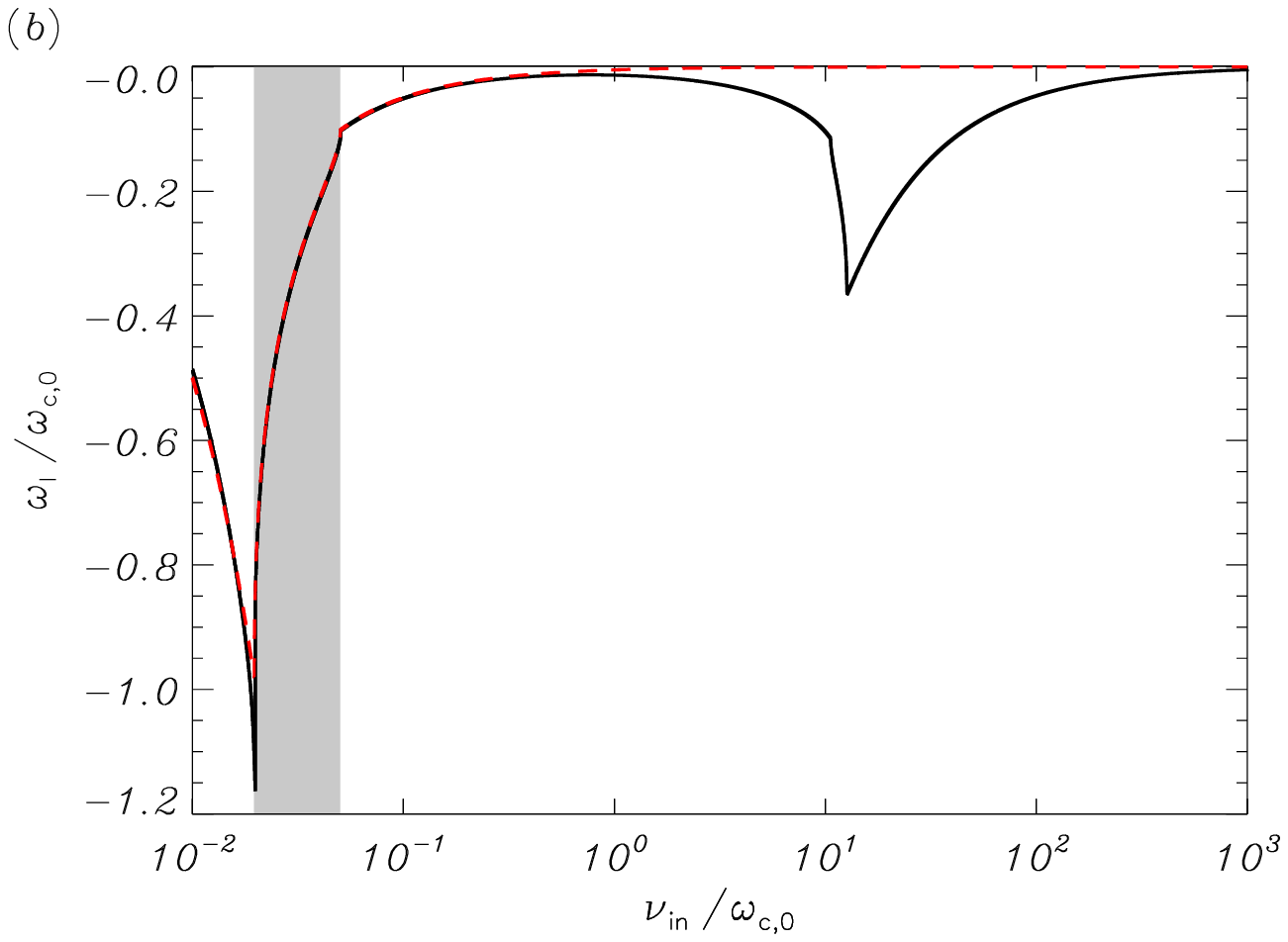}
\caption{Same as Figure~\ref{fig:slow2} but with $\chi=100$. \label{fig:slow3}}
\end{figure}

Now we consider $\chi > 8$ and repeat the previous computations (see Figure~\ref{fig:slow2} for $\chi=20$). As in the case without neutral pressure, we notice the presence of a cut-off region that agrees well with the range given by Equation~(\ref{eq:rangenoprop}). The existence of the cut-off region is not affected by neutral pressure, i.e., the location of the cut-off region in the same when neutral pressure is neglected. For the set of parameters considered in Figure~\ref{fig:slow2} the cut-off region appears in the vicinity  of $\nuin / \omega_{\rm c,0} \approx 10^{-1}$. We increase the ionization fraction to $\chi = 100$ (Figure~\ref{fig:slow3}).  In addition to the cut-off region described above, now we find a second cut-off region. The second cut-off region appears for relatively high collision frequencies,  $\nuin / \omega_{\rm c,0} \approx 10^{1}$ for the set of parameters used in Figure~\ref{fig:slow3}.  This new cut-off region is only present when neutral pressure is taken into account. The result in the absence of neutral pressure completely misses this second cut-off region. The existence of a second cut-off region seems a result of slow modes only, since it is not found in the case of transverse kink modes. Unfortunately, unlike the first cut-off region we do not have an analytic expression that gives us the location of the second cut-off region. Instead, out numerical study informs us that the second cut-off appears when $\chi \gtrsim 24$, although this value may be affected by the particular choice of parameters used in the model.

As for the  transverse kink modes, it is worth mentioning that some mechanisms not considered in the present analysis may produce  damping of the slow mode comparable or even stronger than that due to neutral-ion collisions. This is the case of non-adiabatic mechanisms, e.g., thermal conduction and radiative losses \citep[see, e.g.,][]{demoortel2003,soler2008}. Partial ionization increases the efficiency of thermal conduction due to the additional contribution of neutral conduction \citep[see][]{forteza2008,soler2010}, so that the combined effect of thermal conduction and partial ionization produces a very efficient damping of the slow mode \citep[see, e.g.,][]{khodachenko2006,solerphd}.

\section{Discussion}
\label{sec:conc}

In this paper we have explored the impact of partial ionization on the properties of the MHD waves in a cylindrical magnetic flux tube using the two-fluid formalism. Unlike previous works \citep[e.g.,][]{kumar2003,soler2012} we have considered a consistent description of the dynamics of the neutral fluid that takes gas pressure into account. We have derived the dispersion relation for the wave modes  that is the two-fluid generalization of the dispersion relation of \citet{edwin1983}. Rather than performing particular applications, we have considered the neutral-ion collision frequency as an arbitrary parameter. The modifications of the wave frequencies due to neutral-ion collisions have been explored.

First, we have investigated the case that neglects neutral pressure. This case was previously investigated for surface waves in a Cartesian interface by \citet{kumar2003}.  For $\nuin / |\omega| \ll 1$ we have recovered the frequencies in fully ionized plasma \citep{edwin1983}. As the ratio $\nuin / |\omega|$ increases, both ion-electrons and neutrals become more and more coupled. For $\nuin / |\omega| \gg 1$ ion-electrons and neutrals behave as a single fluid. Compared to the fully ionized case, when $\nuin / |\omega| \gg 1$ the wave frequencies are lower and depend on the ionization ratio, $\chi$. The frequencies are reduced by the factor $\left(\chi+1\right)^{-1/2}$ compared to their values in the fully ionized case. This result is the same for all wave modes and is equivalent to replace the density of ions by the sum of densities of ions and neutrals.  The same conclusion was reached in the previous work by \citet{kumar2003}. Concerning the damping by neutral-ion collisions, we find that damping is weak when $\nuin / |\omega| \ll 1$ unless the plasma is very weakly ionized, i.e., $\chi \gg 1$. Damping is most efficient when the wave frequency is of the same order of magnitude as the collision frequency. Damping is again weak  when $\nuin / |\omega| \gg 1$. This last result is always true regardless the value of $\chi$. Again, our results agree well with the previous findings of \citet{kumar2003}.

Then we have incorporated the effect of neutral gas pressure. Since transverse kink waves are largely insensitive to the sound velocity, we have obtained the same results as in the absence of gas pressure. However, the consideration of neutral pressure has dramatic consequences for slow magnetoacoustic modes.  For high collision frequencies the slow mode frequency is larger than the value obtained in the absence of neutral pressure. This is so because the slow mode frequency depends on an effective sound velocity that corresponds to the weighted average of the sound velocities of ions-electrons and neutrals. When $\chi \gg 1$ the effective sound velocity tends to the sound velocity of neutrals. Concerning the imaginary part of the frequency, the results with and without neutral pressure also show significant differences for high collision frequencies. The imaginary part of the slow mode frequency has an additional minimum at high collision frequencies when neutral pressure is included. This causes the slow mode to be more efficiently damped by collisions when neutral pressure is taken into account.

 Here we go back to the discussion given in the Introduction about the necessity of using the multi-fluid theory instead of the simpler single-fluid MHD approximation. In the solar atmosphere the expected value of the neutral-ion collision frequency \citep[see][Figures~1 and 2]{depontieu2001} is much higher than the frequency of the observed waves  \citep[e.g.,][]{depontieu2007,okamoto2011}. In such a case, the results of this paper point out that, for practical purposes, it is enough to use the single-fluid MHD approximation, but taking into account the following simple recipe  to adapt the results of fully ionized models to the partially ionized case. First of all, the ion density has to be replaced by the total density, i.e., the sum of densities of ions and neutrals. Second, the effective sound velocity should be used instead of the sound velocity of the ionized fluid. This effective sound velocity is the weighted average of the sound velocities of ions-electrons and neutrals. This approach is appropriate for those works which are not interested in the details of the interaction between ions and neutrals, but only on the result of this interaction on the wave frequencies. Hence, this recipe may be useful to perform magneto-seismology of chromospheric spicules and other partially ionized structures as prominence threads. For example, in their seismological analysis of transverse kink waves in spicules \citet{verth2011} considered a definition of the kink velocity that incorporates the density of neutrals. Although our recipe provides a good approximation to obtain the wave frequencies, it misses the effect of damping due to neutral-ion collisions, which might have an impact on the seismological estimates using the amplitude of the waves. Taking into account the range of observed wave frequencies, the damping of kink waves due to neutral-ion collisions may be negligible compared to other effects as, e.g, resonant absorption \citep{soler2012}. However, the damping of slow magnetoacoustic waves may be important even for high collision frequencies. Other damping effects as, e.g., resonant absorption and non-adiabatic mechanisms, should be  considered in addition to neutral-ion collisions to explain the damping of MHD waves in the solar atmosphere \citep[see, e.g.,][]{khodachenko2006,soler2012}. 

On the other hand, a relevant result obtained in this paper  is the presence of cut-offs for certain combinations of parameters \citep[see][]{kulsrud1969}. This result cannot be obtained in the single-fluid MHD approximation. At the cut-offs the real part of the frequency vanishes, meaning that perturbations are evanescent in time instead of oscillatory. For transverse kink waves we find a cut-off region for low values of $\nuin / \omega_{\rm k}$ when $\chi >8$. For slow magnetoacoustic waves there are two cut-off regions. The first one is the same as that found for transverse kink waves, whereas the second one is only present in the case of slow modes when  neutral pressure is considered. The second cut-off region takes place when $\chi \gtrsim 24$ and relatively high collision frequencies. The relevance of these cut-offs for wave propagation in the solar atmosphere should be investigated in forthcoming works.

Finally, we are aware that the model used in this paper is simple and misses effects that may be of importance in reality. The  model should be improved in the future by considering additional ingredients not included in the present analysis. Among these effects, the variation of physical parameters, e.g., density, ionization degree, etc., along the magnetic wave guide, ionization and recombination processes, and the influence of a dynamic background are worth being explored in the near future.

\begin{acknowledgements}
We acknowledge the anonymous referee for his/her constructive comments. We thank T. V. Zaqarashvili for reading an early draft of this paper and for giving helpful comments. RS, JLB, and MG acknowledge support from MINECO and FEDER funds through project AYA2011-22846. RS and JLB  acknowledge support from CAIB through the `Grups Competitius' scheme and FEDER funds. MG acknowledges support from KU Leuven via GOA/2009-009. AJD acknowledges support from MINECO through project AYA2010-1802.
\end{acknowledgements}

\bibliography{refs}

\begin{appendix}

\section{Expression of the dispersion relation}
\label{app}

The dispersion relation is $\mathcal{D}_m \left( \omega, k_z \right)  = 0 $, with $\mathcal{D}_m \left( \omega, k_z \right)$ given by the solution of the following determinant,
\begin{equation}
\mathcal{D}_m \left( \omega, k_z \right)  = \left|  
\begin{array}{cccc}
 a_{11} & a_{12}  & a_{13} & a_{14}  \\
 a_{21} & a_{22}  & a_{23} & a_{24}  \\
 a_{31} & a_{32}  & a_{33} & a_{34}  \\
 a_{41} & a_{42}  & a_{43} & a_{44}  
\end{array}
\right|,  \label{eq:genreldisper}
\end{equation}
with
\begin{eqnarray}
 a_{11} &=& a_{12} = a_{13} = a_{14} =1, \\
a_{21} &=& \frac{q_{2,0}^2}{k_{1,0}^2 - k_{\rm n,0}^2}, \qquad a_{22} = \frac{q_{2,0}^2}{k_{2,0}^2 - k_{\rm n,0}^2}, \\
 a_{23} &=& \frac{q_{\rm 2,ex}^2}{k_{\rm 1,ex}^2 - k_{\rm n,ex}^2}, \qquad a_{24} = \frac{q_{\rm 2,ex}^2}{k_{\rm 2,ex}^2 - k_{\rm n,ex}^2}, 
 \end{eqnarray}
\begin{eqnarray}
a_{31} &=& \frac{k_{\rm 1,0}}{\rho_{\rm i,0}\left( \tilde{\omega}^2 - \omega_{\rm A,0}^2 \right)} \left( 1 - \frac{i \nuin}{\omega + i \nuin} \frac{q_{2,0}^2}{k_{1,0}^2 - k_{\rm n,0}^2}\right) \nonumber \\ &\times &\frac{J'_m(k_{\rm 1,0} R)}{J_m(k_{\rm 1,0} R)}, 
\end{eqnarray}
\begin{eqnarray}
a_{32} &=& \frac{k_{\rm 2,0}}{\rho_{\rm i,0}\left( \tilde{\omega}^2 - \omega_{\rm A,0}^2 \right)} \left( 1 - \frac{i \nuin}{\omega + i \nuin} \frac{q_{2,0}^2}{k_{2,0}^2 - k_{\rm n,0}^2}\right) \nonumber \\ &\times &\frac{J'_m(k_{\rm 2,0} R)}{J_m(k_{\rm 2,0} R)}, 
\end{eqnarray}
\begin{eqnarray}
a_{33} &=& \frac{k_{\rm 1,ex}}{\rho_{\rm i,ex}\left( \tilde{\omega}^2 - \omega_{\rm A,ex}^2 \right)} \left( 1 - \frac{i \nuin}{\omega + i \nuin} \frac{q_{\rm 2,ex}^2}{k_{\rm 1,ex}^2 - k_{\rm n,ex}^2}\right) \nonumber \\ &\times & \frac{H^{'(1)}_m(k_{\rm 1,ex} R)}{H^{(1)}_m(k_{\rm 1,ex} R)}, 
\end{eqnarray}
\begin{eqnarray}
a_{34} &=& \frac{k_{\rm 2,ex}}{\rho_{\rm i,ex}\left( \tilde{\omega}^2 - \omega_{\rm A,ex}^2 \right)} \left( 1 - \frac{i \nuin}{\omega + i \nuin} \frac{q_{\rm 2,ex}^2}{k_{\rm 2,ex}^2 - k_{\rm n,ex}^2}\right) \nonumber \\ &\times & \frac{H^{'(1)}_m(k_{\rm 2,ex} R)}{H^{(1)}_m(k_{\rm 2,ex} R)},
\end{eqnarray}
\begin{eqnarray}
a_{41} &=& k_{\rm 1,0} \left[ \frac{i \nuin}{\omega + i \nuin}\frac{1}{\rho_{\rm i,0} \left( \tilde{\omega}^2 - \omega_{\rm A,0}^2 \right)} \right. \nonumber \\ &\times & \left( 1 - \frac{i \nuin}{\omega + i \nuin}\frac{q_{2,0}^2}{k_{1,0}^2 - k_{\rm n,0}^2}  \right) \nonumber \\
&-& \left. \frac{1}{\rho_{\rm n,0} \omega \left( \omega + i \nuin \right)}\frac{q_{2,0}^2}{k_{1,0}^2 - k_{\rm n,0}^2} \right] \frac{J'_m(k_{\rm 1,0} R)}{J_m(k_{\rm 1,0} R)}, 
\end{eqnarray}
\begin{eqnarray}
a_{42} &=& k_{\rm 2,0} \left[ \frac{i \nuin}{\omega + i \nuin}\frac{1}{\rho_{\rm i,0}\left( \tilde{\omega}^2 - \omega_{\rm A,0}^2 \right)} \right. \nonumber \\ &\times & \left( 1 - \frac{i \nuin}{\omega + i \nuin}\frac{q_{2,0}^2}{k_{2,0}^2 - k_{\rm n,0}^2}  \right)  \nonumber \\
&-& \left. \frac{1}{\rho_{\rm n,0} \omega \left( \omega + i \nuin \right)}\frac{q_{2,0}^2}{k_{2,0}^2 - k_{\rm n,0}^2} \right] \frac{J'_m(k_{\rm 2,0} R)}{J_m(k_{\rm 2,0} R)}, 
\end{eqnarray}
\begin{eqnarray}
a_{43} &=& k_{\rm 1,ex} \left[ \frac{i \nuin}{\omega + i \nuin}\frac{1}{\rho_{\rm i,ex}\left( \tilde{\omega}^2 - \omega_{\rm A,ex}^2 \right)} \right. \nonumber \\ &\times & \left( 1 - \frac{i \nuin}{\omega + i \nuin}\frac{q_{\rm 2,ex}^2}{k_{\rm 1,ex}^2 - k_{\rm n,ex}^2}  \right)  \nonumber \\
&-& \left. \frac{1}{\rho_{\rm n,ex} \omega \left( \omega + i \nuin \right)}\frac{q_{\rm 2,ex}^2}{k_{\rm 1,ex}^2 - k_{\rm n,ex}^2} \right] \frac{H^{'(1)}_m(k_{\rm 1,ex} R)}{H^{(1)}_m(k_{\rm 1,ex} R)}, 
\end{eqnarray}
\begin{eqnarray}
a_{44} &=& k_{\rm 2,ex} \left[ \frac{i \nuin}{\omega + i \nuin}\frac{1}{\rho_{\rm i,ex}\left( \tilde{\omega}^2 - \omega_{\rm A,ex}^2 \right)} \right. \nonumber \\ &\times & \left( 1 - \frac{i \nuin}{\omega + i \nuin}\frac{q_{\rm 2,ex}^2}{k_{\rm 2,ex}^2 - k_{\rm n,ex}^2}  \right) \nonumber \\
&-& \left. \frac{1}{\rho_{\rm n,ex} \omega \left( \omega + i \nuin \right)}\frac{q_{\rm 2,ex}^2}{k_{\rm 2,ex}^2 - k_{\rm n,ex}^2} \right] \frac{H^{'(1)}_m(k_{\rm 2,ex} R)}{H^{(1)}_m(k_{\rm 2,ex} R)}.
\end{eqnarray}

\end{appendix}

\end{document}